\documentclass[sigconf,screen,nonacm]{acmart}

\AtBeginDocument{%
  \providecommand\BibTeX{{%
    \normalfont B\kern-0.5em{\scshape i\kern-0.25em b}\kern-0.8em\TeX}}}

\acmPrice{15.00}
\acmISBN{978-1-4503-XXXX-X/18/06}

\renewcommand\footnotetextcopyrightpermission[1]{}
\usepackage{graphicx} % Required for inserting images
\usepackage{subcaption}
\usepackage{float}
\usepackage{comment}
\usepackage{tcolorbox}
\usepackage{tipa}
\usepackage{makecell}

\usepackage{titlesec}

\titleformat{\paragraph}[runin]{\normalfont\normalsize\bfseries}{}{0pt}{}
\titlespacing*{\paragraph}{0pt}{*1}{1em}

\newcommand{\visowel}{\textit{\textbf{V}(is)\textbf{owel}}}

\begin{document}

\title[\visowel{}]{\visowel{}: An Interactive Vowel Chart to Understand What Makes Visual Pronunciation Effective in Second Language Learning}
%%
%% The "author" command and its associated commands are used to define
%% the authors and their affiliations.
%% Of note is the shared affiliation of the first two authors, and the
%% "authornote" and "authornotemark" commands
%% used to denote shared contribution to the research.

\settopmatter{authorsperrow=2}
\author{Charlotte Kiesel}
% \authornote{Both authors contributed equally to this research.}
\email{yoder6@illinois.edu}
\affiliation{%
  \institution{Univ. of Illinois in Urbana-Champaign}
  % \city{Champaign}
  % \state{Illinois}
  \country{}
}
\orcid{0009-0006-7316-9351}

\author{Dipayan Mukherjee}
\email{dipayan2@illinois.edu}
\affiliation{%
  \institution{Univ. of Illinois in Urbana-Champaign}
  % \city{Champaign}
  % \state{Illinois}
  \country{}
}
\orcid{0000-0002-6017-1739}

\author{Mark Hasegawa-Johnson}
\email{jhasegaw@illinois.edu}
\affiliation{%
  \institution{Univ. of Illinois in Urbana-Champaign}
  % \city{Champaign}
  % \state{Illinois}
  \country{}
}
\orcid{0000-0002-5631-2893}

\author{Karrie Karahalios}
\email{kkarahal@illinois.edu}
\affiliation{%
  \institution{Univ. of Illinois in Urbana-Champaign}
  % \city{Champaign}
  % \state{Illinois}
  \country{}
}
\orcid{0000-0001-8788-3405}

%%
%% By default, the full list of authors will be used in the page
%% headers. Often, this list is too long, and will overlap
%% other information printed in the page headers. This command allows
%% the author to define a more concise list
%% of authors' names for this purpose.
\renewcommand{\shortauthors}{Kiesel et al.}

\begin{abstract}
%topic under study and imply the underlying question. One sentence
%Two sentences: previous research has demonstrated. Provide rationale for new research
%Two sentences: the data, research, and analytical methods used in my study
% major findings and implications and significance of this study
Visual feedback speeds up learners' improvement of pronunciation in a second language. The visual combined with audio allows speakers to see sounds and differences in pronunciation that they are unable to hear. Prior studies have tested different visual methods for improving pronunciation, however, we do not have conclusive understanding of what aspects of the visualizations contributed to improvements. Based on previous work, we created \visowel{}, an interactive vowel chart. Vowel charts provide actionable feedback by directly mapping physical tongue movement onto a chart. We compared \visowel{} with an auditory-only method to explore how learners parse visual and auditory feedback to understand how and why visual feedback is effective for pronunciation improvement. The findings suggest that designers should include explicit anatomical feedback that directly maps onto physical movement for phonetically untrained learners. Furthermore, visual feedback has the potential to motivate more practice since all eight of the participants cited using the visuals as a goal with \visowel{} versus relying on their own judgment with audio alone. Their statements are backed up by all participants practicing words with \visowel{} more than with audio-only. Our results indicate that \visowel{} is effective at providing actionable feedback, demonstrating the potential of visual feedback methods in second language learning.
\end{abstract}


%%
%% Keywords. The author(s) should pick words that accurately describe
%% the work being presented. Separate the keywords with commas.

% \received{}
% \received[revised]{12 March 2009}
% \received[accepted]{5 June 2009}

\maketitle
% To ponder: What is the goal of this paper? What would I want to know if I was reading it? What is the narrative? What makes this important research?
\section{Introduction}
% Motivation
% Pronunciation is crucial in the acquisition of a second language (L2). While pronunciation can gradually improve during acquisition through implicit immersion, explicit pronunciation practice improves it at a faster rate~\cite{carey_call_2004,hew_effect_2004,offerman_visual_2016,yuen_enunciate_2011,sherwin_natural_2018,skalski_mapping_2011,vanden2013more}. Learners who practice pronunciation with visual feedback outperform their peers who used other methods of practice~\cite{carey_call_2004,hew_effect_2004,offerman_visual_2016,yuen_enunciate_2011}. Many visual feedback systems have a steep learning curve to understand and act on the feedback. 
% While many studies interview learners about their experience post interaction with a tool, there is lack of understanding of how interactions with visual versus audio-only feedback affects learners awareness of their pronunciation as a whole. 

% why is learning pronunciation in a second language is hard
Every speaker comes with their own language background and physical characteristics, thus, pronunciation feedback must be personalized for second language learners. Individualizing feedback requires more time from language teachers in a system already strained by the lack of instructors~\cite{moser2024covid,hismanoglu2010language,rao2019key,munro2015setting}. Computer-Assisted Pronunciation Training (CAPT) has the potential to assist learners by providing feedback specific to their needs. Although CAPT introduces the potential for algorithmic error and requires access to computing resources, it addresses the drawbacks of a teacher-dependent model. It does not require extra time from a teacher, is widely distributable, and provides access to practice in a self-paced environment. Among CAPT methods, visual feedback has arisen as an effective way to provide pronunciation feedback~\cite{carey_call_2004,hew_effect_2004,offerman_visual_2016,yuen_enunciate_2011}.   

% why do we need to study visual feedback methods? What has been done? What is lacking?
There are two main types of visual feedback for pronunciation: correctness indicators, such as color coding or percentages, and articulation-based representation of speech. Many well-known language learning apps, such as Duolingo, babbel, and Rosetta Stone, use correctness-based visual feedback, while previous research focuses on articulation-based visualizations (ABV)~\cite{carey_call_2004,hew_effect_2004,offerman_visual_2016}. We choose an ABV because it satisfies all the criteria in Bliss et. al.~\cite{bliss2018computer} for effective feedback. Correctness-based feedback fails two of the criteria; namely, feedback must be (i) natural and logical and (ii) able to facilitate comparison~\cite{bliss2018computer}. Since color coding and percentages do not specify what about the production is incorrect, correctness-based feedback fails to be natural and logical. It fails the latter by forcing learners to use their ears to determine the differences instead of visually representing comparisons. 

% , such as simple color coded levels of correct and incorrect,
% 
\begin{figure*}[t]
    \centering
    \begin{tabular}{c c}  % Two-column image layout
        \begin{subfigure}{0.45\textwidth}
            \centering
            \includegraphics[width=\linewidth]{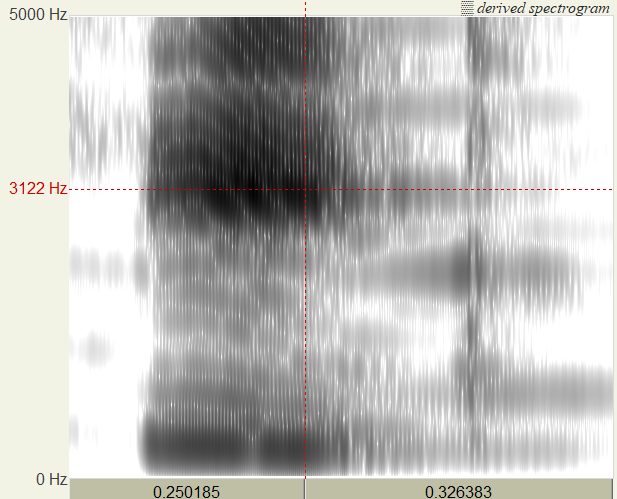}
            \caption{A spectrogram from Praat of the English word "Heat".}
            \label{fig:specHeat}
        \end{subfigure} &
        \begin{subfigure}{0.45\textwidth}
            \centering
            \includegraphics[width=\linewidth]{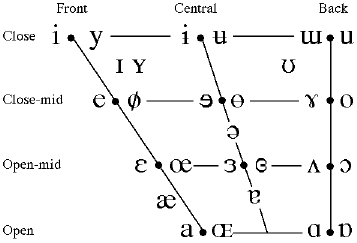}
            \caption{An example vowel chart~\cite{wieling2011inducing}.  The x-axis represents the second formant and frontedness or backedness of the tongue. The y-axis is the first formant and height of the tongue in the mouth.}
             \label{fig:exampleVwlChart}
        \end{subfigure} \\

        \begin{subfigure}{0.45\textwidth}
            \centering
            \includegraphics[width=\linewidth]{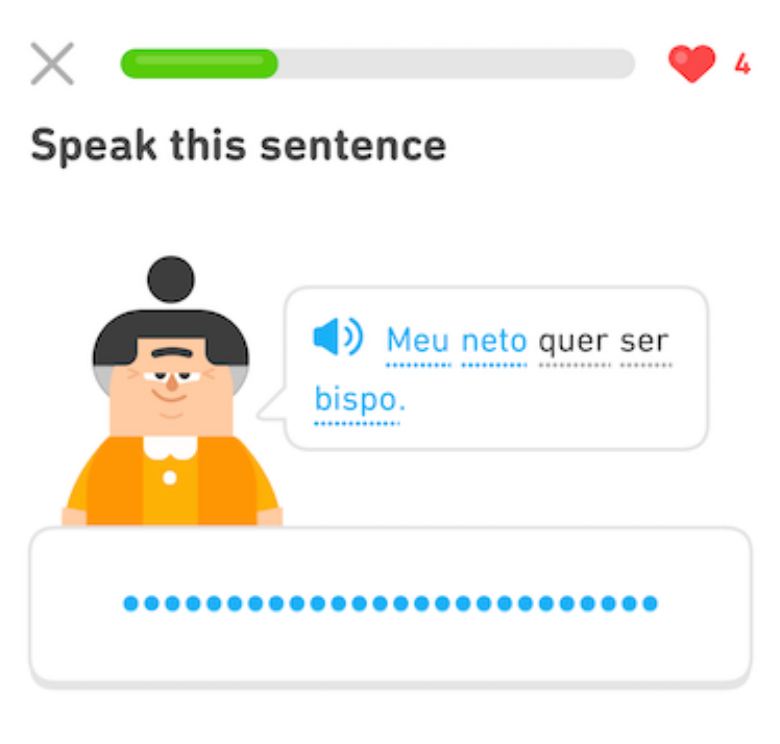}
        \caption{A screenshot of Duolingo's binary speaking feedback as of 2020~\cite{Blanco_Moline_2025}}.
        \label{fig:VOTexample}
        \end{subfigure} &
        \begin{subfigure}{0.45\textwidth}
            \centering
            \includegraphics[width=\linewidth]{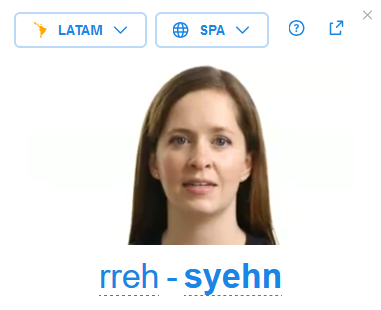}
        \caption{A still frame from a video of a speaker's face on SpanishDict.com.}
        \label{fig:facialFeature}
        \end{subfigure}
    \end{tabular}

    % \vspace{10pt} % Adjust spacing before the caption

    \caption{Four types of pronunciation visualizations.}
    \label{fig:proVisualizationEx}
\end{figure*}
Three widely studied visual feedback methods are the spectrogram, animated face-cutaways, and the vowel chart. The spectrogram, which can be used for both vowels and consonants, is difficult to interpret without training and hard to modify pronunciation based on its feedback (Figure~\ref{fig:specHeat}). Animated face-cutaways and vowel charts, which provide feedback on a subset of sounds, have fewer parts to interpret and uses the principle of natural mapping to visualizes articulator movement (Figure~\ref{fig:exampleVwlChart}). The vowel chart follows more closely the principles of effective feedback and is effective in improving pronunciation~\cite{rehman2021real,carey_call_2004}. Previous implementation of vowel charts, however, do not fully integrate audio and visual. Users had to adjust pronunciation based either on visual or audio output. For this reason, we created an interactive version, \visowel{}. The chart displays the position of the tongue based on the first and second formants, prominent bands of energy based around a frequency~\cite{schaferFormant1970}. \visowel{} provides broad applicability by accepting vowels within multiple contexts including rhotics, like the American /r/\footnote{The forward slash (/) represents the orthographic character in a language. Square brackets around a character indicate a phonetic description.} and displays a vowel's change over time, allowing learners to train the difference between single vowels (monophthongs) and combined vowels (diphthongs). As far as we are aware, we are the first study to have vowels in multiple contexts, visuals for diphthongs, and \textbf{the first study in visual Computer-Assisted Pronunciation Training (CAPT) to capture how learners interpret visual results as they adjust their pronunciation}.

%A spectrogram visualizes the amplitudes of frequencies in a sound signal. A vowel chart uses the principle of natural mapping which has been applied in many domains with positive results to map the position of the tongue to a position on the chart~\cite{sherwin_natural_2018,skalski_mapping_2011,vanden2013more}.
%obile interface design, physical automobile controls, and video game controller design with positive results. 

% Research Questions

Using an iterative design approach, we created \visowel{} and an audio-only practice tool. We ran a within-subject study with 8 participants and elicited speakers' real-time thought processes as they used both tools. 

\visowel{} encouraged participants to practice more because it gave them a goal to work toward or pushed them to reflect on the first language speaker's production and their own when visual results did not align with their expectations. Based on our findings, we suggest that future pronunciation feedback should include a visual representation of the closeness of the learner and target speaker.

Our work makes three main contributions.
\begin{enumerate}
    \item Learners' thought processes during interaction with visual pronunciation feedback.
    \item An interactive vowel chart, \visowel{}, that can be used in any word context.
    \item A technical implementation for extracting vowels from any context.
\end{enumerate}

%Existing Best Practices
\section{Related Work}
% Why Pronunciation is important
Pronunciation in second language acquisition (SLA) is interconnected with the building blocks of language-- grammar, vocabulary, and listening comprehension-- as well as integration into a language community. Thus, improving pronunciation leads to improvement in other areas of SLA and vice versa~\cite{sicola2015integrating,derwing2003esl}.  There are multiple ways that a first language (L1) of a speaker, usually a language spoken from a very young age, can interfere with acquiring sounds in a second language (L2). Sounds that distinguish meaning in an L1, called phonemes, may not distinguish meaning in an L2 or vice versa. As a result, interlocutors can be confused when communicating with an L2 speaker~\cite{rao2019key}. We discuss previous work that focuses on human- and computer- assisted methods for pronunciation improvement. 
% as well as work on designing visualizations for audio.  

%Past Research
%Non-tech
\subsection{Human-Mediated Pronunciation Training}
Little attention is given to pronunciation in textbooks and curricula other than at a collegiate level in the United States of America. This makes ad hoc reactive feedback for pronunciation common in foreign language classrooms from K-12~\cite{rao2019key,wei2006literature,darcy2021window}. Learners who receive ad hoc reactive feedback do not improve significantly in comparison with intentional methods of training. As a result, we leave this out of the discussion despite it being a common approach in classrooms~\cite{darcy2021window}. Beyond reactive feedback, three main methods of pronunciation instructor-dependent training are outlined below: reading out loud, mimicry of an L1 speaker, and minimal pair drills (drilling words that differ by one sound/phoneme\footnote{For example, hat/hit, fit/feet, cat/bat are minimal pairs in English. Hat and catch are not since they differ by more than one sound}).  

\subsubsection{Reading Out Loud}
\label{sec:reading}
A commonly studied method of pronunciation improvement is reading out loud~\cite{thomson_effectiveness_2015, chenRetAssist2024}. This can be achieved through paired practice, where peers listen and give feedback, mimicking practice, or group reading, where the whole classroom and teacher listen to the reader and provide feedback on pronunciation. The exercise improves oral reading fluency and reading comprehension~\cite{han2016integrating}, but is dependent on a teacher's input or practice with peers. Because feedback is given in real-time, L2 learners will not receive detailed pronunciation feedback.

\subsubsection{Mimicry of an L1 Speaker}
As mentioned in section \ref{sec:reading}, mimicry can be used in conjunction with reading out loud. It can also be used for shorter segments of pronunciation, such as words or sounds, which is the method we use in our study design. Research shows that mimicry is key in improving pronunciation~\cite{hinton_aptitude_2013, szyszka_good_2015}. While mimicry is used in classrooms, it does not require an active listener. L2 learners can practice mimicry with any kind of medium that includes audio, such as movies or podcasts, though this means they have to rely on their own judgment to adjust pronunciation. 

\subsubsection{Minimal Pair Drills}
\label{sec:minimal}
Minimal pair drills (MPDs) rely on listening and mimicking an L1 speaker to bring out the contrast that different sounds have by putting them in identical contexts. MPDs are used in speech therapy and language learning to great effect though they limit the context that sounds can appear based on the L1~\cite{barlow_minimal_2002, hamzah_teaching_2019, rahman_use_2018, brown_minimal_1995}. Despite potential downsides of minimal pairs, we contend that visually contrasting vowels will benefit the L2 learner.   

% Tech
\subsection{Computer-Assisted Pronunciation Training (CAPT)}
CAPT has the advantage over traditional pronunciation feedback by its ability to be used by many and be automatically catered to the individual. CAPT feedback takes two different forms: audio-only and visual accompanied by audio. The latter outperforms traditional and audio-only feedback~\cite{olson2014benefits,hew_effect_2004}. Visual feedback falls into two main categories: spectrograms and articulatory-based visualizations.

\subsubsection{Spectrograms} 
Spectrograms are useful for their flexibility in pronunciation feedback. They can be used to train vowels and consonants including the effects those sounds have on each other (Figure \ref{fig:specHeat}). Studies have shown the positive impact on pronunciation when students reflect on their speech with the aid of a spectrogram. But a spectrogram can require hours of training to correctly interpret and previous studies use it as a static tool instead of an interactive one~\cite{greene_recognition_1984}, likely because it is difficult to apply feedback. Using it as a static tool limits learners to reflect on their pronunciation without immediately adjusting their pronunciation. 

\subsubsection{Articulatory-based Visualizations}
Articulatory-based visualizations, on the other hand, are used in phonetic training to teach learners about articulators, such as the tongue and lips, in many different languages. Previous work focuses on two ABV: face-cutaway animations and vowel charts.

Animated face-cutaway diagrams display how articulators move during speech. Learners can view correct movement based on the animation and attempt to mimic the example speaker~\cite{bu2021pteacher}. However, this work does not display an animation of a learner's incorrect speech or how to change pronunciation from incorrect to be correct~\cite{bu2021pteacher}. As a result, feedback targets mistakes but does not provide feedback to move from current pronunciation to the target pronunciation.

Vowel charts display a vowel's location on a trapezoid. There are two ways to represent vowels on a vowel chart -- with IPA symbols (Figure \ref{fig:exampleVwlChart}) or with formants (Figure \ref{fig:freqVwlChart}). Previous studies show that vowel charts are effective within limited contexts in improving pronunciation ~\cite{paganus2006vowel,rehman2021real}. One method studied vowels by themselves while another limited it to vowel preceded by an 'h' and followed by a 'd'~\cite{paganus2006vowel,rehman2021real}. Both restrictions limit the real-world use of the tool. Furthermore, how learners understand what they see on an interactive vowel chart is unclear since all the studies interview participants in a reflexive way, making it difficult to know how to improve the interaction experience to facilitate a better understanding of a vowel chart. Finally, these vowel charts do not integrate the audio with the visual, requiring learners to listen and interpret visuals separately~\cite{paganus2006vowel,rehman2021real}.  

We designed a real-time interactive vowel chart system, \visowel{}, that would function for vowels in any context and provide integrated audio and visual feedback. To understand why the visualization is effective, we held think-aloud experiments and compared how learners think about visual feedback in contrast to audio-only feedback. By understanding how untrained learners interpret visual feedback, we can modify existing visualizations to promote improvement and minimize confusion.

\begin{figure*}[ht]
    \centering
    \includegraphics[width=0.8\linewidth]{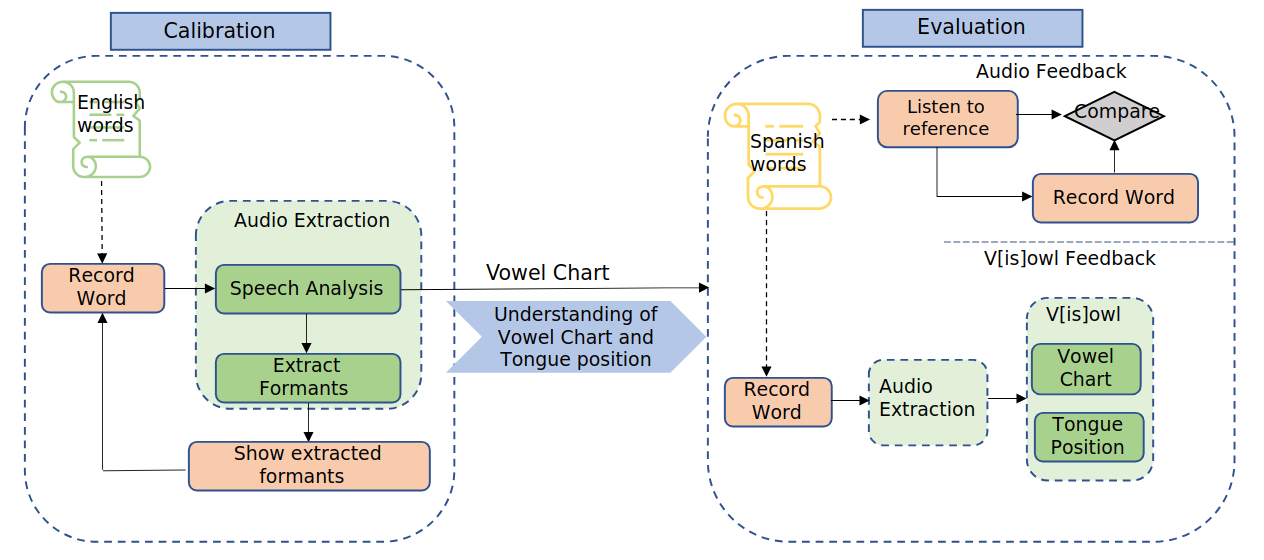}
    \caption{The study begins by calibrating a user's voice using extracted formants of four English vowels [\textipa{i},u,a,\textipa{6}]. After calibration, users interact with an audio-only feedback system and \visowel{}, which extracts formants and displays them on a vowel chart. }
    \label{fig:workflowPrac}
\end{figure*}

% \subsection{Visual Feedback for Audio Signals}
% \todo{write this section :)}

\subsection{Research Questions}
As we seek to understand what makes a visualization effective, we need to make sure we view it holistically. If a visualization is effective in improving the target sounds, such as vowels, but does not affect or negatively affects overall pronunciation, we should consider visualizations that communicate information for more sounds. In short, to understand what makes visualizations effective and if they limit speakers during practice, we seek to answer the following questions:

\begin{itemize}
    \item[\textbf{RQ1:}] How does interaction between a visual and audio-only feedback method differ?
    \item[\textbf{RQ2:}] How do phonetically untrained users interpret vowel charts during interaction?
    \item[\textbf{RQ3:}] How does a visualization that focuses on vowels affect users' perceptions of other aspects of pronunciation?
\end{itemize}
% Because it is not clear what aspects of visualizations learners find useful during practice, we ask questions that help us know how users interpret the visual feedback. 
% Research indicates that higher motivation in language learning correlates with commitment to learning a language; the longer one learns a language the better the pronunciation becomes.

% In general, studies show that immediate explicit pronunciation feedback is beneficial for L2 learners~\cite{darcy2021window,lacabex_explicit_2020,yakut_promoting_2020, sepasdar_immediate_2019}.

% \footnote{A monophthong is one sound and a diphthong is the blending of two sounds, for example, "I" in English is generally said as a diphthong.}
% \input{2design}
\section{Design}
In this section, we outline the design of our interactive vowel chart, \visowel{}. We discuss the interface of \visowel{} and its calibration and tutorial, then we present our design of \visowel{}'s signal processing backend.

\subsection{User Interface}
\visowel{} is comprised of a vowel chart, which displays the Spanish speaker's and user's vowels, recording buttons, and audio playback button for the Spanish speaker. When any vowel is clicked, including the Spanish speaker's, audio plays. In order for \visowel{} to display two speakers on one chart, the user must first have their voice calibrated. Once their voice is calibrated, users are able to complete a tutorial introducing them to \visowel{} where they practice with English words and Spanish words to begin to build mental models of how it works. 

\begin{figure}[t]
    \centering
    \begin{subfigure}{\linewidth}
        \centering
        \includegraphics[width=\linewidth]{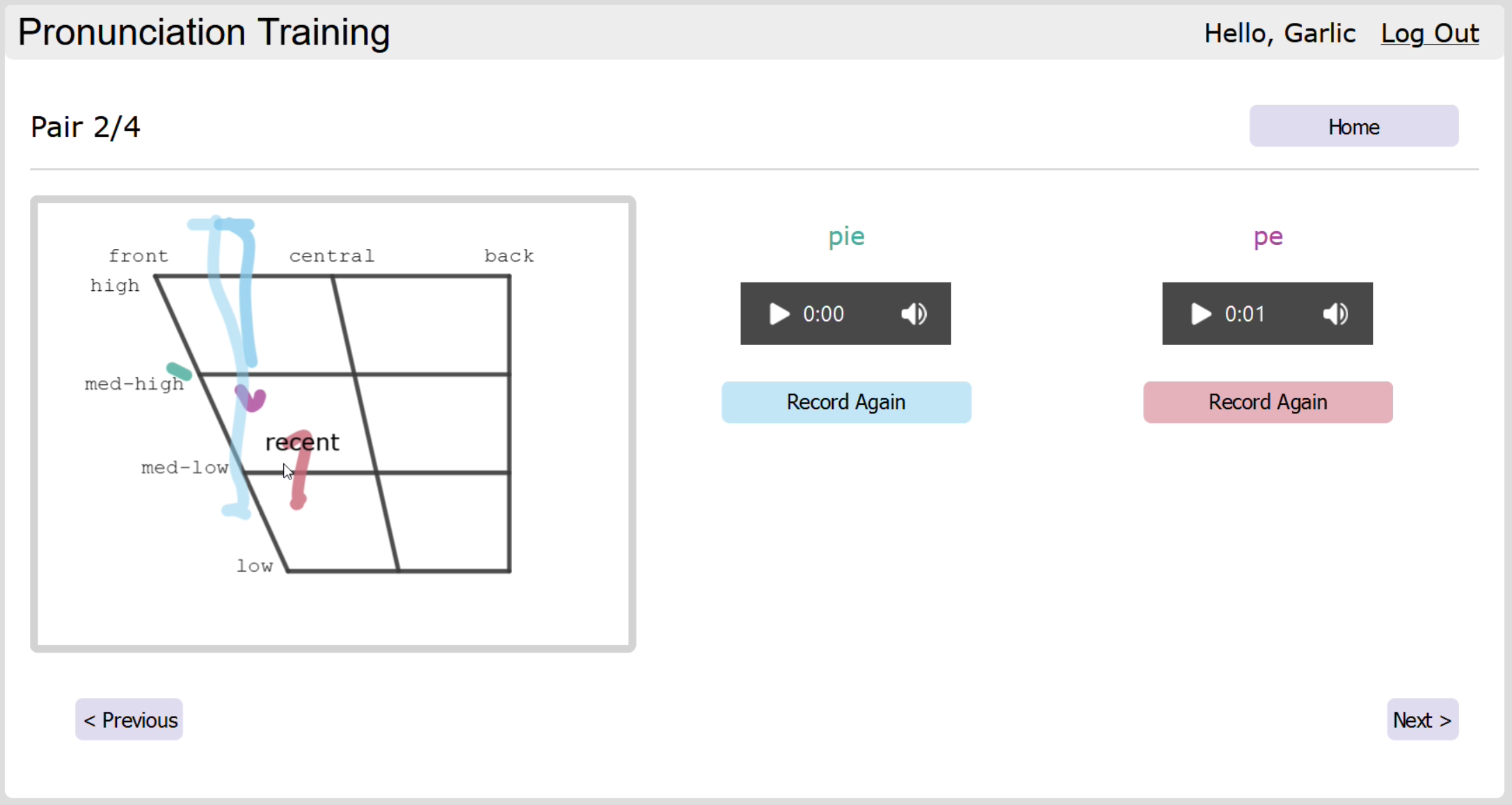}
        \caption{A practice page for \visowel{}. The user has recorded twice for "pie" and once for "pe". The most recent recording is "pe". }
        \label{fig:visowelPrac}
    \end{subfigure}
    \newline
    \begin{subfigure}{\linewidth}
        \centering
        \includegraphics[width=\linewidth]{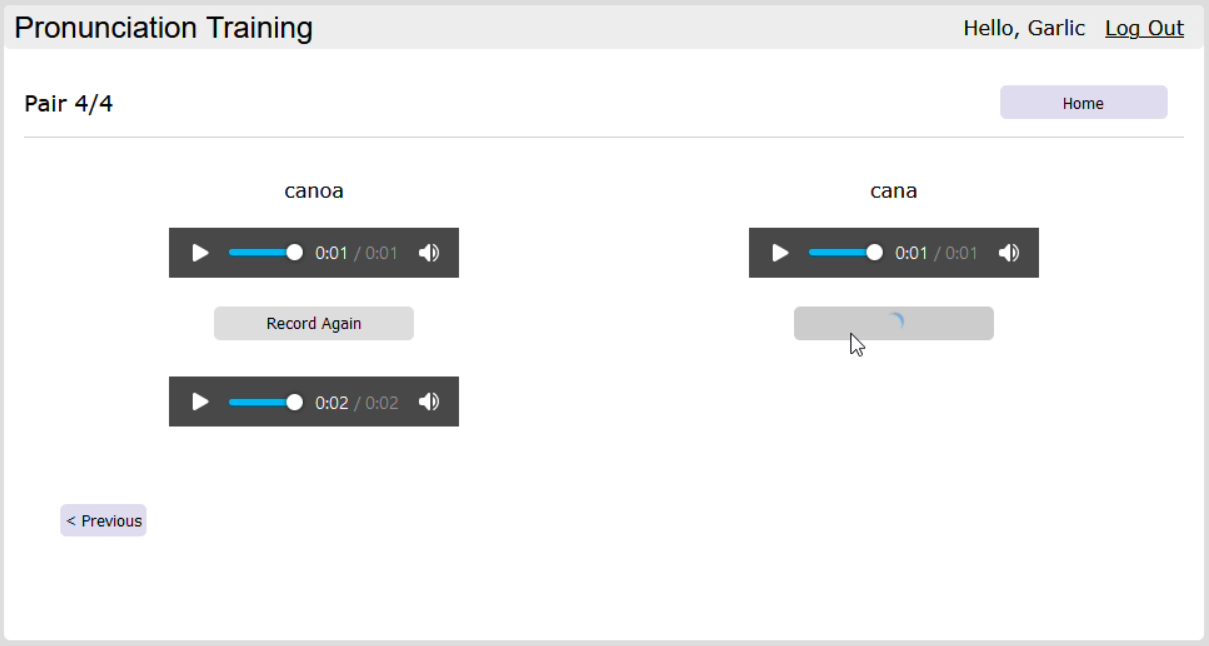}
        \caption{A practice page for audio-only. The user is waiting to record "cana", as shown by the semi-circle in the record button.}
        \label{fig:audioPrac}
    \end{subfigure}
    
    \caption{Example practice pages from audio-only and \visowel{}.}
    \label{fig:pracPages}
\end{figure}

\begin{figure}
    \centering
    \includegraphics[width=1\linewidth]{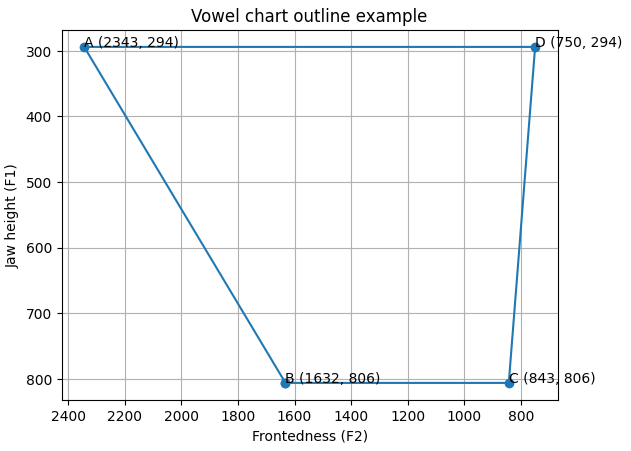}
    \caption{A vowel chart with first and second formant frequency markings.}
    \label{fig:freqVwlChart}
\end{figure}
% Calibration first
% 
\subsubsection{\visowel{}} 
The final design of \visowel{} is trapezoidal shaped and has descriptive labels on the top and left sides, e.g. front, mid, back, etc. Both of these design decisions were inspired by a traditional vowel chart (Figure \ref{fig:exampleVwlChart}). 

\begin{figure}[t]
    \centering
    \includegraphics[width=0.85\linewidth]{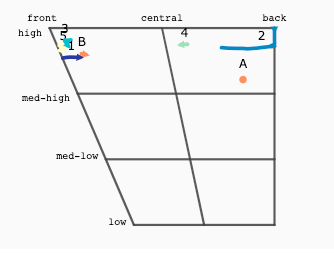}
    \caption{\visowel{} with Spanish vowels marked by A and B and user's vowels labeled numerically with the latest recording starting at 1.}
    \label{fig:manyRecordings}
\end{figure}

\paragraph{Vowel lines.} Vowels are plotted on the chart based on their change in frequencies over time. Clicking on a plotted vowel plays the associated audio for the extracted vowel then the whole word. If the vowel moved over time, an animation redraws the line starting from the beginning. Initially, we attached a simple arrow on the end of the line to indicate vowel change direction (Figure \ref{fig:manyRecordings}). The arrowhead helped, but due to changes in formants, it would sometimes point the wrong direction on the line. In the final version, we removed the arrowhead and added an animation of the line when it was clicked to communicate change over time. 

The vowel trajectory is based on averaging neighboring formants. We began by directly plotting all extracted formant values but since extracted formants vary slightly, this created noisy lines. Because the difference in neighboring formants is due to slightly inaccurate formant extraction and not a speaker's tongue movement, we decreased the visual noise by averaging neighboring formants. 

\begin{figure}[t]
    \centering
    \includegraphics[width=0.65\linewidth]{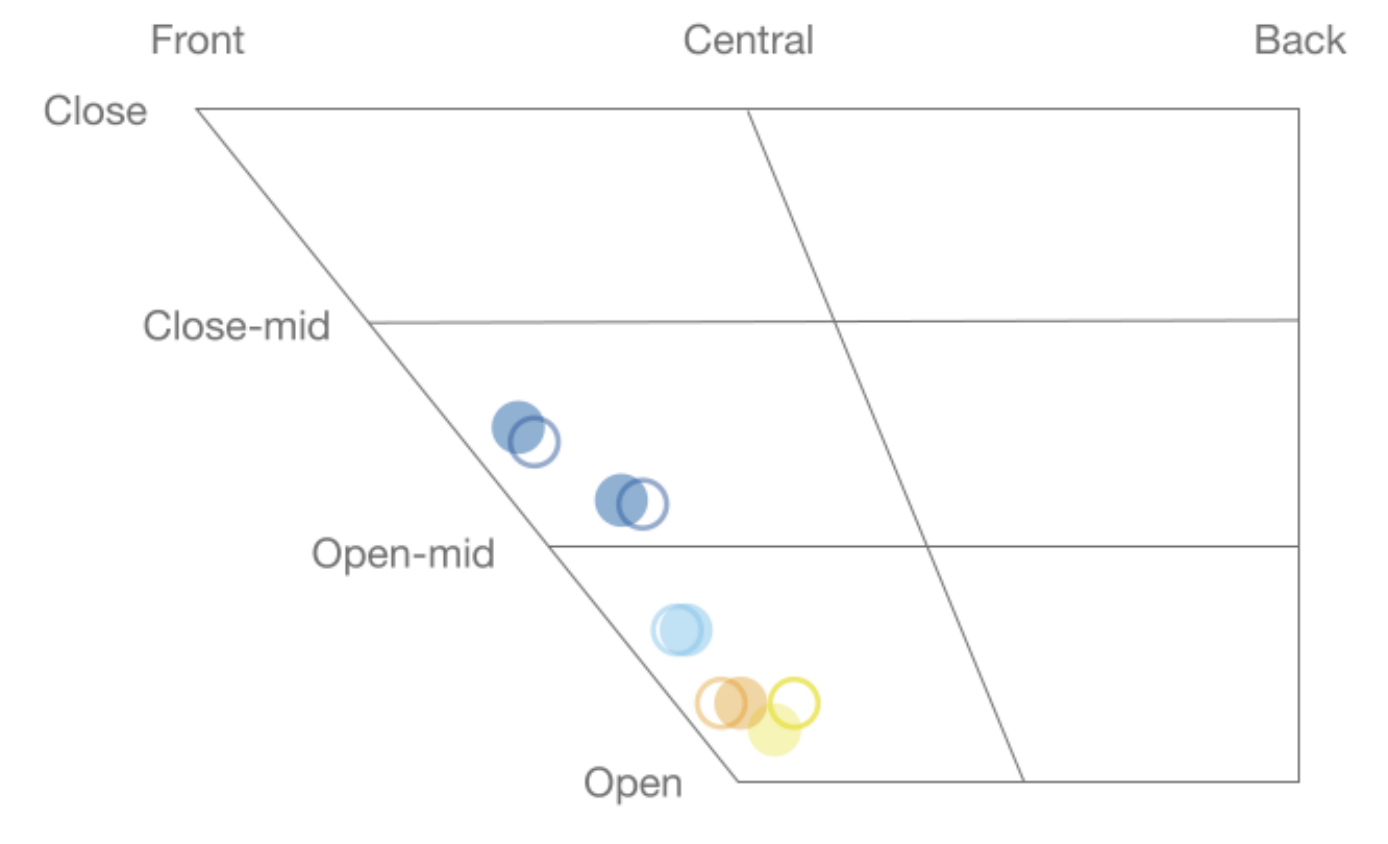}
    \caption{A mock-up of two vowels from one word plotted on a vowel chart. The first vowel's dot is colored in and the second vowel's dot is outlined. }
    \label{fig:twoVowels}
\end{figure}

\paragraph{Multiple recordings.} A user can record words as many times as they would like, however, only the most ten most recent recordings are displayed on the chart. Each time a new vowel is plotted on the chart, the previous vowels become more transparent. In the beginning phases of design, we did not limit the number of vowels on the chart nor the number of vowels extracted from the word (Figure \ref{fig:twoVowels}). The mock-up tested well, but practically, plotting even one vowel proved visually overwhelming when the vowel was long, did not stay in the same place, or was recorded more than a few times. Many recordings created visual clutter on the chart, so we limited the total number of recordings to ten per word and use opacity to communicate the ordering of recordings.

\paragraph{Grid lines and labels.} In our first design, we followed previous work and marked frequencies along the axes (Figure \ref{fig:freqVwlChart})~\cite{rehman2021real,paganus2006vowel}. Discussion with the pretest group showed that people were unsure how to interpret vowels marked on the chart. Our research team decided to strip a vowel chart from all labels, then reintroduce labels one at a time to gauge understanding of the chart based on available labels. Frequency labels did not allow group members to understand what that meant with respect to their physical movements while physical labels (front, back, high, low) were more helpful in interpretation.

Although \visowel{} has bounding grid lines, vowels are not constrained to them. When we forced vowels within the trapezoid, crucial information was lost regarding the current tongue position with respect to previous recordings (Figure \ref{fig:manyRecordings}). For example, the vowel "ea" ([i]\footnote{Square brackets around a character indicate a phonetic description.}) in the English word "beat", could be plotted as slightly above or further forward than the previous recordings, which would put it outside the chart. Removing this information would inaccurately represent the tongue's position as the same between two recordings even though it had moved. As a result, we let vowels be plotted outside the chart. We kept vowels within a wider box around the grid lines, so that even if our extraction algorithm failed to pick out the correct frequencies for the formant, users would have visual feedback.

% Once recording starts, the interface displays a spinning circle for 0.5 seconds to allow the speech recognition algorithm to establish a noise threshold, then recording bars replace the circle. Once the speaker has stopped speaking, the algorithm extracts the vowel boundaries (onset and offset times) and formants of the vowel. Initially, we used written feedback, \textit{"Please wait..."}, that would appear after a user pressed record. Pretests revealed that users would immediately speak as soon as they pressed "Record", only later realizing that the text asked them to wait. When we switched to the spinning symbol, further tests showed that the change caused the users to pause until recording bars appeared. 

\begin{figure}[t]
    \centering
    \includegraphics[width=0.45\textwidth]{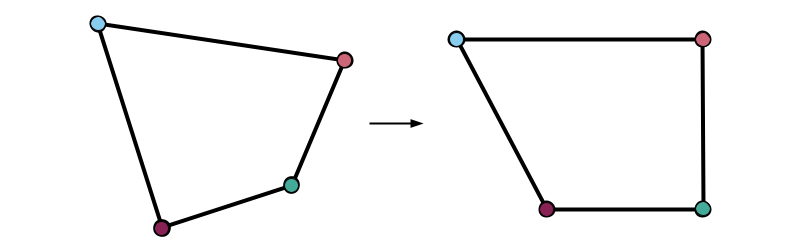}
    \caption{On the left is an example frequency shaped trapezoid using corner vowels. We use a projective transformation to map  the observed formant frequencies and to a consistent vowel chart representation.}
    \label{fig:transform}
\end{figure}
% \paragraph{\textbf{Calibration}}
% Before a user can interact with \visowel{}, their voice is transformed from the frequency domain into the SVG domain. The transformation comes from a learner's elicitation of the English "corner" vowels, [i,u,\ae,\textturnscripta(6)] extracted from the words beet, boot, bat, and bought. By basing each transformation on a specific speaker's vowels, visualizations more closely represent human perception of sounds by removing speaker-unique characteristics and retaining language-dependent features (Figure \ref{fig:transform}).

\paragraph{\textbf{Calibration}} User voices can vary significantly depending on the individual. This variety will change the shape of the vowel space per individual, as shown by the left image in Figure \ref{fig:transform}. We transform the individual vowel space into a fixed trapezoid, as shown by the right image in Figure \ref{fig:transform}, which removes idiosyncratic speech characteristics and retains language-dependent features. We collect the user's specific vowel space using their elicitation of the English "corner" vowels, [i,u,\ae,\textturnscripta(6)] extracted from the words beet, boot, bat, and bought. We use projective transformation (homography) to map the user-specific frequencies to our vowel chart. Since there is no set ratio for vowel chart trapezoids, we chose the trapezoid ratio for height, top width, and bottom width as 1, 1.33, and 0.833 respectively. The height represents the spread of the first formant ($f_1$) frequency, and the width represents the spread of the second format at different $f_1$. This shape is consistent with the known spread of the vowel chart. The consistent shape of the vowel chart enhances the aesthetics of the chart and the interpretability of results across users.

%% Talk about the mathematics of the process

\paragraph{\textbf{Tutorial}} 
The tutorial for \visowel{} leads learners through understanding a vowel chart. It introduces the tongue's position for producing vowels, how to understand the axes of the vowel chart, and finally, how to mimic a speaker based on their plotted vowel. We settled on this flow based on pretest user interaction with paper and interactive prototypes (details in Appendix \ref{app:tutorial}). We begin with informing users of the tongue's role in producing a vowel, specifically focusing on closeness to the roof of the mouth, tongue height, and where it is in the mouth, close to the teeth or away. We started with the tongue's position, since pretest users were not sure how the tongue's position related to vowel production. After an overview of the chart in relation to tongue position, we give them opportunities to focus on pairs of English vowels that differ in tongue position and height. By starting with known vowels, we were able to bake interaction into the examples and avoid a lot of written instruction, since pretest users often skipped reading.

\begin{figure}[h]
    \centering
    \begin{subfigure}{\linewidth}
        \centering
        \includegraphics[width=0.9\linewidth]{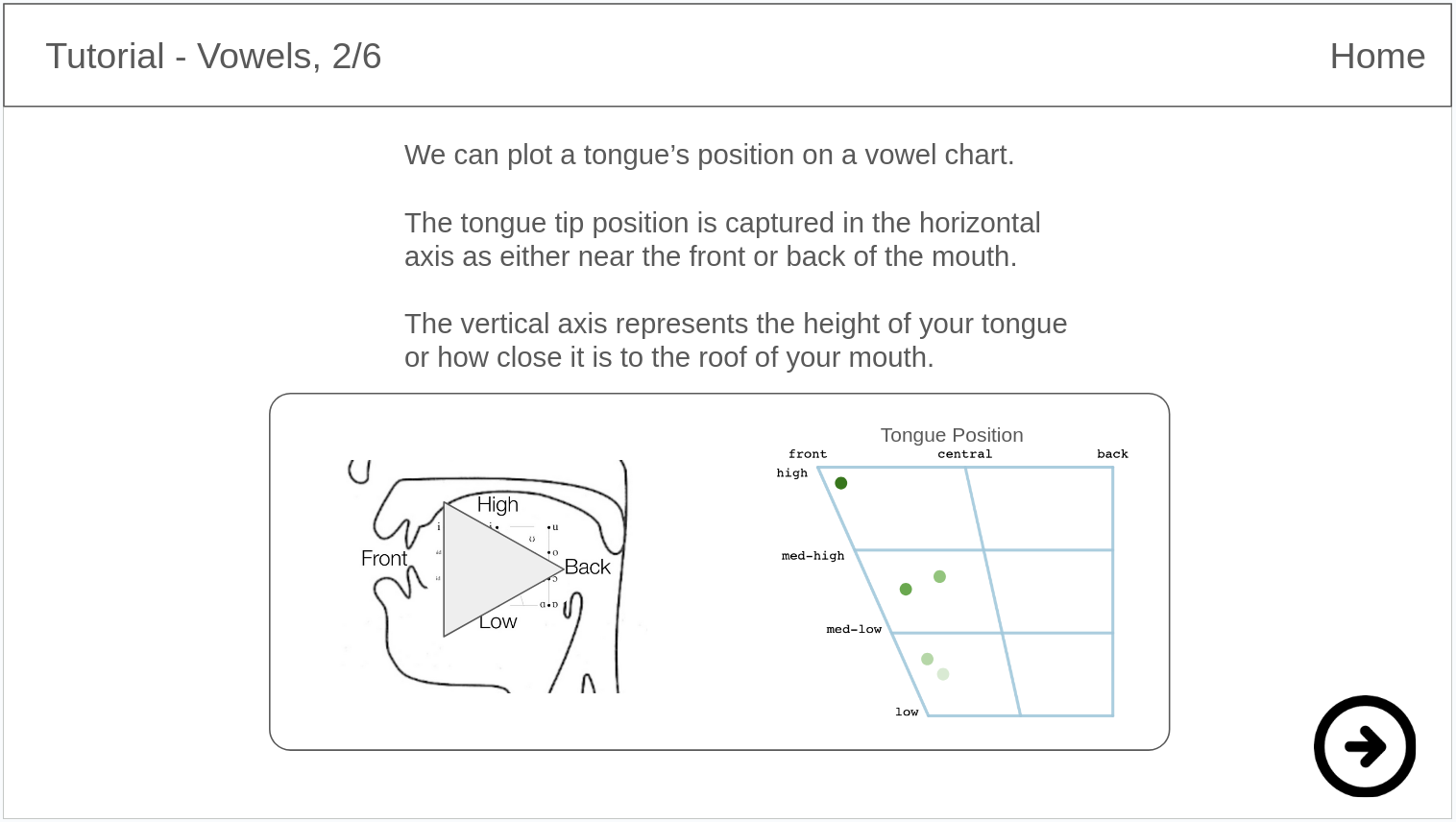}
        \caption{The second page of the paper prototype. One face cutaway is on the left and on the right a vowel chart has multiple different vowels plotted on it.}
        \label{fig:2tutPaper}
    \end{subfigure}
    \begin{subfigure}{\linewidth}
        \centering
      \includegraphics[width=0.825\linewidth]{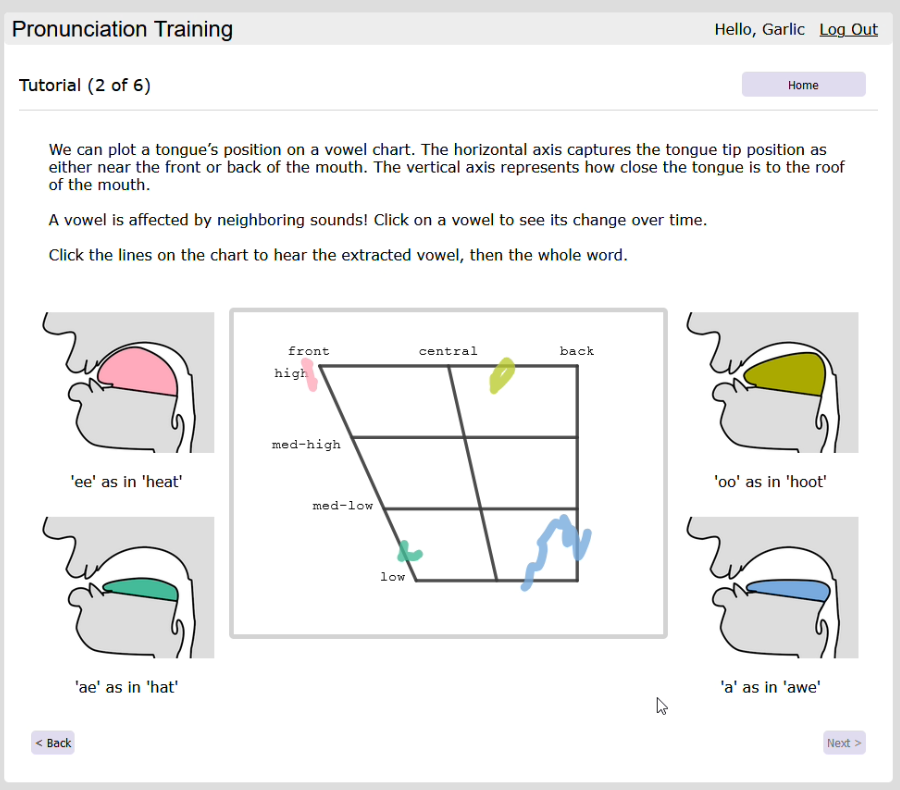}
        \caption{Final interface for the tutorial's second page. The vowels on the chart share the same color as the tongue in their respective face cutaway.}
        \label{fig:2tut}
    \end{subfigure}
    \label{fig:pg2}
    \caption{Design cycle of page 2}
\end{figure}

\subsection{Technical Backend}
\begin{figure*}[ht]
    \centering
    \includegraphics[width=\linewidth]{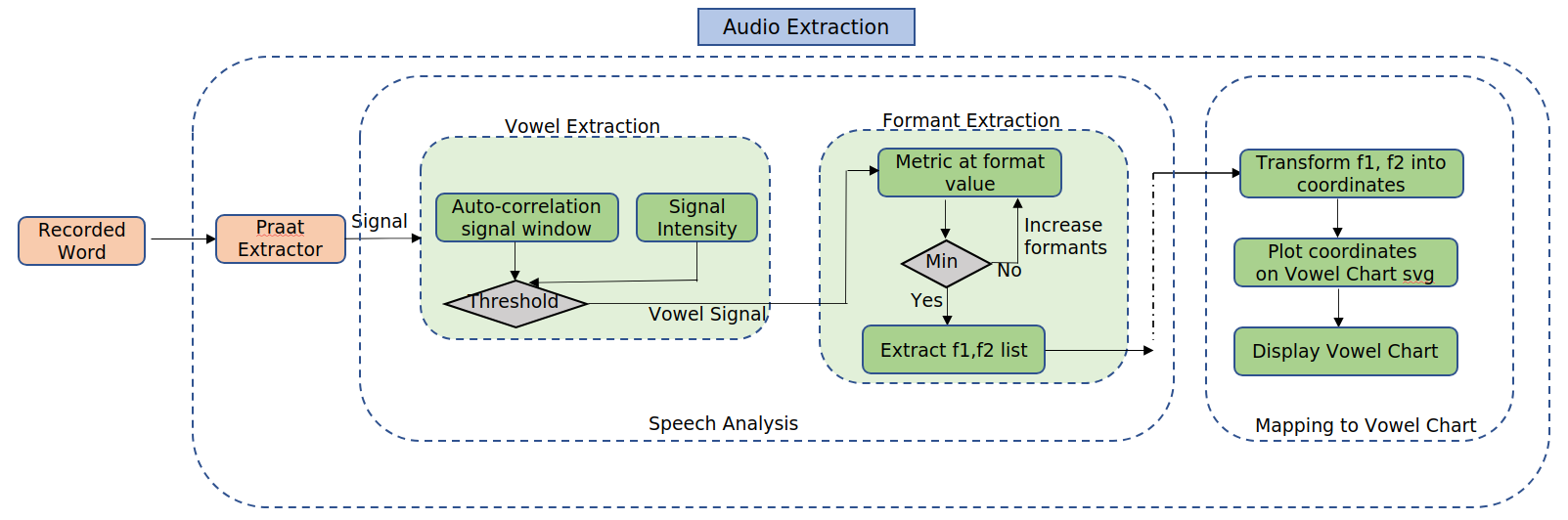}
    \caption{A flowchart representing the algorithmic process of extracting vowels and formants. }
    \label{fig:extractAudio}
\end{figure*}
We discuss the design of the signal processing algorithm and its accuracy based on pre-study results. The algorithm extracts the vowel and vowel formants, then transforms the formants into the html viewbox according to a speaker's transformation (Figure~\ref{fig:extractAudio}). An audio file with a single word is passed as a signal to the vowel extractor. The vowel extraction uses the autocorrelation of the signal to determine the onset and offset times of the vowel. A ".wav" file is created using these timestamps. The new file is passed into formant extraction which tests different numbers of formants and calculates a goodness metric. The algorithm extracts $f_1$ and $f_2$ values using the formant with the best metric. The lists of $f_1$ and $f_2$ are transformed into the viewbox then displayed on the vowel chart. 

\subsubsection{Vowel Extraction}
\begin{figure}[t]
    \centering
    \includegraphics[width=\linewidth]{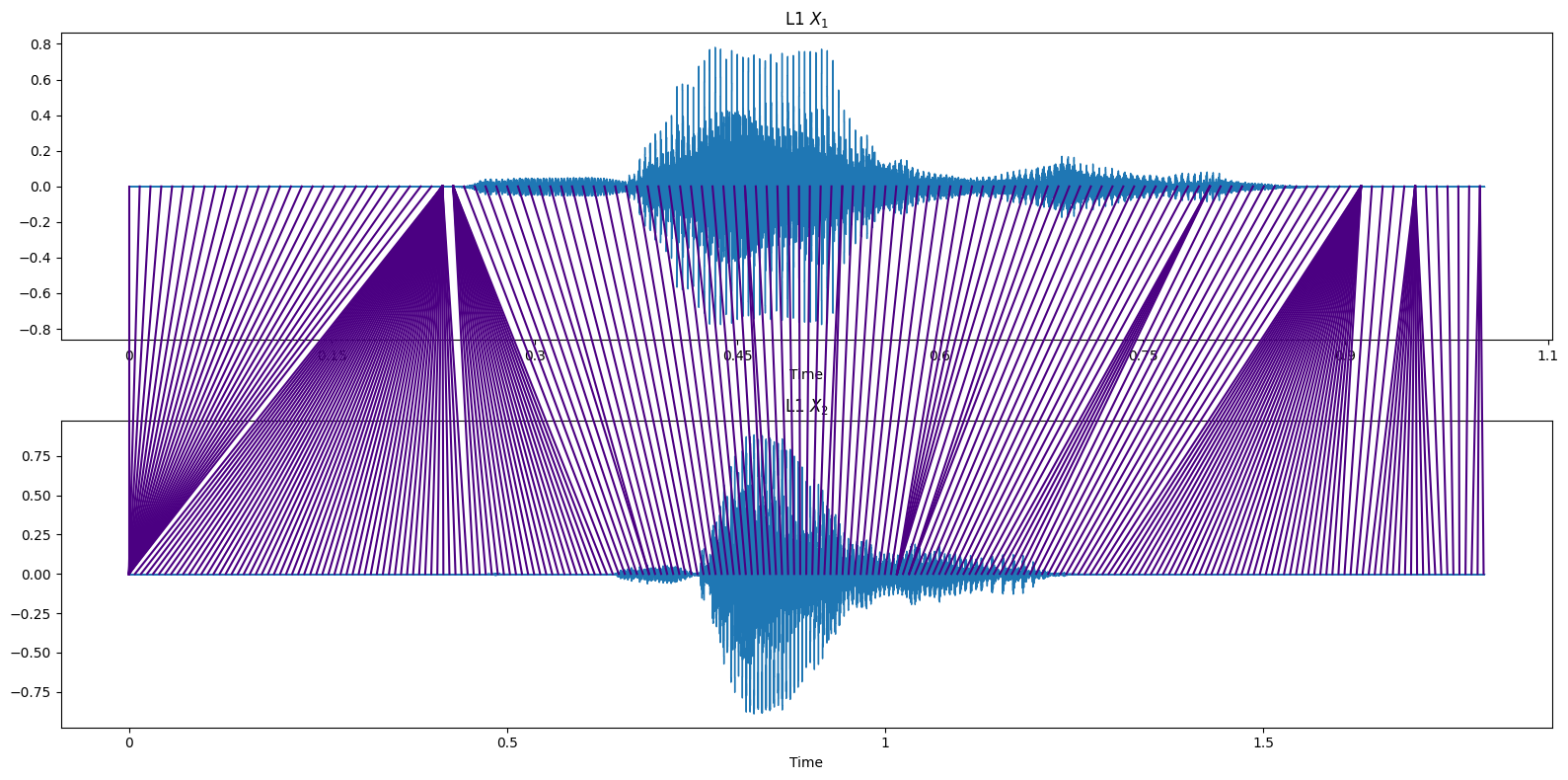}
    \caption{An example graph we made demonstrating the Dynamic Time Warping (DTW) algorithm~\cite{bellman1959dtw} which matches similar parts of two different signals. The purple lines represent the connections DTW made.}
    \label{fig:DTW}
\end{figure}
We designed a vowel extraction algorithm based on the autocorrelation of the signal after testing two existing algorithms that did not have the precision needed, Dynamic Time Warping (DTW)~\cite{bellman1959dtw} and PocketSphinx Phoneme Aligner (PS)~\cite{huggins2006pocketsphinx}, as shown in Table \ref{tab:algoTimes}. We tested the algorithms on 5 voices from our lab, 3 female. Even though DTW generally had a lower absolute difference, it tended to mark the beginning of the vowel during the consonant, as represented by the negative. We determined the precision needed through trial and error. Starting 20 milliseconds after the expected beginning of the vowel will not miss diphthong information but starting more than 10ms before the start of the vowel will pick up on too much of the preceding consonant vice versa for vowel offset. 

\begin{figure}[ht]
    \centering
    \includegraphics[width=\linewidth]{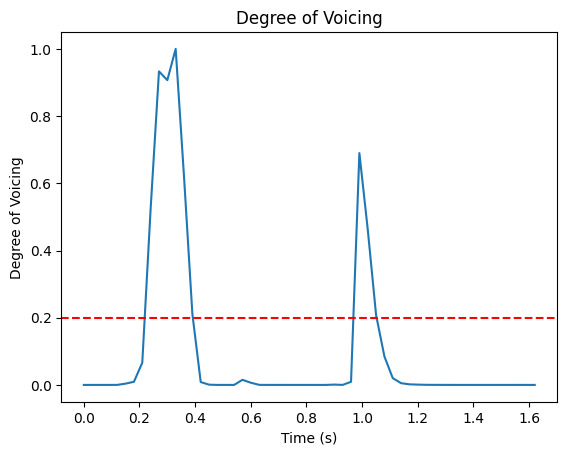}
    \caption{An example degree of autocorrelation for the word 'bata' of a signal with the threshold marked horizontally in red. The two peaks are the vowels.}
    \label{fig:extractAudio}
\end{figure}

Our novel signal processing algorithm extracts the beginning and end of the vowel by calculating the degree of autocorrelation within a sliding window, where autocorrelation is computed as follows:

\begin{align*}
    c_k = \sum_n v_{n+k} \cdot \overline{v}_n
\end{align*}

where $v$ is the audio signal and $\overline{v}_n$ is the complex conjugate.  
Once the autocorrelation goes over a predefined threshold and then back under, the code marks the timestamps (Figure \ref{fig:extractAudio}). We set the threshold by experimenting on different words and speakers to find the degree of autocorrelation needed to be considered a vowel. This algorithm produced the most accurate timestamps for the beginning of the vowel though it still produced inaccurate endings based on whether the following consonant was also fairly resonant. 

\begin{table}[!ht]
    \centering
    \begin{tabular}{ll}
        Algorithm & Average diff. to Manual Calculation \\ \hline
        DTW & -8.2ms \\ 
        sphinx & 25.4ms \\ 
        Ours & 13.4ms \\ 
    \end{tabular}
    \caption{The average difference in onset times for three vowel extraction algorithms. We ended with ours because it was better to lag than lead the beginning of the vowel}
    \label{tab:algoTimes}
\end{table}

% \subsubsection{formant extraction}

% \todo{Dipayan:}
% We used Praat's formant extraction which requires a frequency ceiling and number of formants. Because the number of formants in a signal changes based on word not speaker, we tested formants extracted using 4, 5, and 6 formants. 
%  \begin{align*}
%      metric &= std (value) \\
%      out method &= do this \label{eq:1}
%  \end{align*}

% We measured the standard deviation (std) across the frequencies and the degree of overlap between the first and second formants. The lower std and lower degree of overlap that a formant has the greater the likelihood that the algorithm found the correct frequencies for the first and second formant. 

\subsubsection{Formant Extraction.}
We use Praat's ~\cite{hugoQuenePraat} format extraction library to retrieve the first two formants. Praat's method requires a frequency ceiling and the number of formants.  We set the frequency ceiling at 5500 Hz. However, since the number of formants in a signal changes based on the word, not the speaker, we designed a metric for a signal, described below, to measure the best formant for the signal. We need the first two formants to be smooth because when the algorithm incorrectly chooses a formant, the formant tends to jump multiple frequencies between points.

 \begin{align*}
     &\textit{smoothness}(sig) = stddev(sig) \\
     &metric(sig,fmt\_cnt) =  \textit{smoothness}(sig_{f1}) + \textit{smoothness}(sig_{f2})\\ &\quad \quad \quad \quad \quad \quad \quad \quad \quad + \sum_{i=3}^{fmt\_cnt} 0.8^{i-2} \textit{smoothness}(sig_{fi})
 \end{align*}

Based on this metric, we choose the formant number which has the lowest metric value. It ensures that the first two formants are consistent for the signal, and the higher formants are also somewhat stable. The stability of the formant ensures that our algorithm has found the correct frequency for the first and second formant.
\section{User Study Methodology}
\begin{figure*}[ht]
    \centering
    \includegraphics[width=0.8\linewidth]{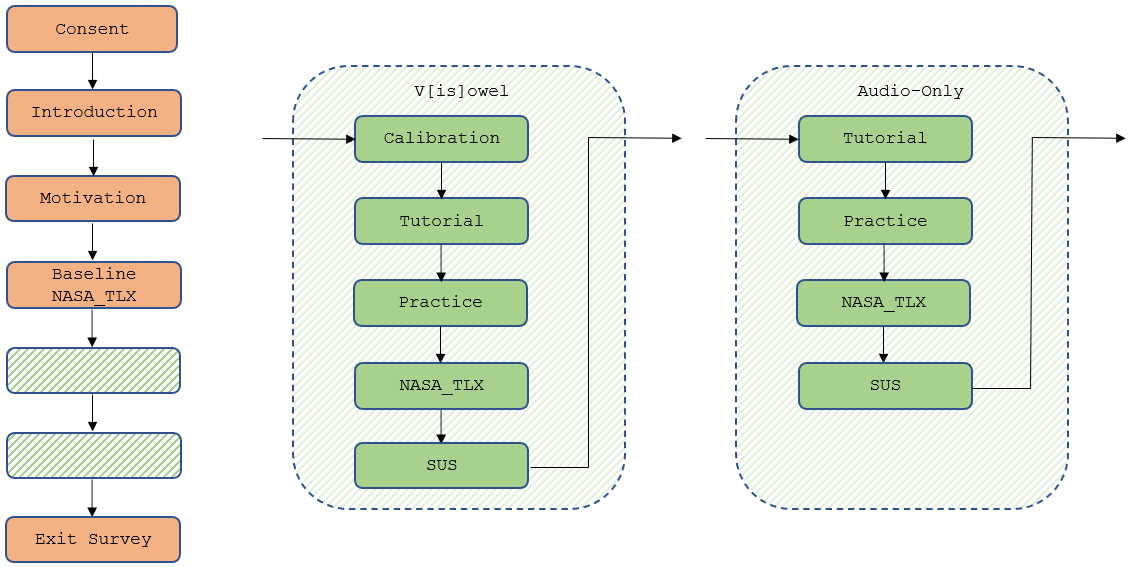}
    \caption{A flowchart representing the progress of the study. The orange boxes represent the fixed portions of the study, the two green striped boxes are the counterbalanced pronunciation practice tools, \visowel{} and audio-only.}
    \label{fig:studyFlow}
\end{figure*}

We held a within-subject study with 8 participants at a large Midwestern university in the United States of America during December of 2024 (Figure \ref{fig:studyFlow}). In this section, we describe the process of recruitment and human-subject evaluation for \visowel{} and audio-only pronunciation practice. We ran two pilot sessions before our study, thus participant numbers start from 2.

\subsection{Recruitment}
\begin{table}[t]
    \centering
    \begin{tabular}{l|lllll|}
        ID  & Word List  & 1st Tool & 2nd Tool & Sex \\ \hline
        % P1 & first & Vowel & Control & Male \\ 
        P2 & first & Control & Vowel & Male \\ 
        P3 & second & Vowel & Control & Female \\ 
        P4 & second & Control & Vowel & Female \\ 
        P5 & first & Control & Vowel & Male \\ 
        P6 & second & Control & Vowel & Male \\ 
        P7 & first & Vowel & Control & Female \\ 
        P8 & second & Vowel & Control & Male \\
        P9 & second & Vowel & Control & Female
    \end{tabular}
    \caption{Participant information}
    \label{tab:participants}
\end{table}

We recruited participants through the campus e-newsletter, and physical posters throughout campus buildings (See Table \ref{tab:participants}). The participants were between the ages of 18 and 24, had no history of speech or auditory impairment, and spoke American English as their first language. We limited participants to those who had not spent more than a semester in college learning Spanish, a week studying Spanish in a Spanish-speaking country, or two months in a Spanish-speaking country. We asked participants to choose the sex of the Spanish speaker with which they wished to practice.

\subsection{Interface Evaluation}

% study design
To understand how participants process their pronunciation during practice with and without a visualization, we conducted one-hour think-aloud experiments in-person on campus. During the practice portion of the experiment, we often prompted them to express why they decided to record a second time. We randomized the order in which participants saw the two practices, \visowel{} and audio-only (Figure \ref{fig:pracPages}). Since we wanted to test the robustness of our signal processing algorithm and visualization, participants practiced with words that included vowels in varied linguistic contexts. These lists were also counterbalanced to appear with both visualizations. 

% we did not have time to 
% In order to see if either visualization had an effect on pronunciation or perception, participants completed 3 pronunciation and perception evaluations of Spanish words. Before, between, and after interaction with \visowel{} and audio-only feedback.

\subsection{Experimental Setup}
Figure \ref{fig:studyFlow} shows the flow of our study. We collected consent, gave a brief introduction to the project, and had participants record English words for a baseline task load (NASA-TLX). Depending on which condition they saw first, the next step was interaction with \visowel{} or audio-only. Audio-only's interface was identical to \visowel{} without a vowel chart (Figure \ref{fig:audioPrac}). The tutorial for audio-only was a spoken introduction to its mimicry setup. Other than an initial calibration step, interactions with the tools followed the same steps: a tutorial introducing the practice, practice with the tool with four different Spanish minimal pairs, followed by a NASA-TLX and System Usability Score (SUS) surveys. After interacting with both tools, we held a semi-structured exit interview. 

% Once participants practice four different minimal pairs, they fill out the NASA-TLX and SUS surveys, which measure the task load and system usability respectively~\cite{hart1986nasa}. The former measures how laborious practicing pronunciation is while the latter helps explain if that labor is due to the difficulty of processing audio and visual data or to the general usability of the system. 

We ran the tools locally on a Lenovo laptop with a Bluetooth microphone recording at 48kHz. The microphone was set on a \~6" stand. Participants were compensated for their time and effort by a \$20 Amazon gift card. They viewed the interface either on a monitor or laptop screen and used a mouse to navigate the interface. 

After explaining and signing the consent form, we briefly introduced the goal of the project. We used NASA TLX to measure workload~\cite{hart1986nasa}, which measures the workload based on mental, physical, and temporal demand as well as performance, effort, and frustration while doing a task. We took a baseline workload assessment to mitigate potential effects of participant, meeting time, location, and screen difference. Participants were asked to separately record three English words as the baseline assessment. We used the process as a baseline since it would show how difficult participants found it to record and speak on our platform without the mingled effect of practicing pronunciation.  
Then, we introduced the tools to the participants and asked that they let us know if they had any questions before they began. The audio-only introduction was verbal, while the introduction to \visowel{} began with calibration and a 6-step tutorial. After the tutorial, they took a 3-question vowel-chart literacy check. We used the check to allow participants to apply what they'd learned and ask clarification questions about the visualization.

We wrapped up the experiment with a semi-structured survey about their experience using both pronunciation practice tools.

\subsection{Data Analysis}
One person on the research team transcribed the interviews and coded the discussions. In the first round of coding, they marked statements as either relating to pronunciation, visual information, auditory feedback, or reaction to the system. During the second round, they broke up the main codes into sub-codes, which are presented in results. We discovered two main themes during analysis of the coded interview data: the role of \textbf{auditory feedback} in understanding speech with and without visual feedback and how \textbf{visual feedback} can help and confuse learners in their goal to improve pronunciation. 
\section{Results}
To answer our research questions, we must look at (i) the difference in thought processes between \visowel{} and audio-only practices, (ii) how and what users considered during interpretation of the vowel chart, and finally, (iii) the effects on participants' focus on pronunciation as a whole during \visowel{} interaction. 

With this in mind, we present our results in two parts: first, with the themes drawn from the qualitative analysis of interaction and interview data then present quantitative findings; second, the technical evaluation of our signal processing algorithm. Quotes from participants are edited for clarity.

Our qualitative results start with participants' reactions to audio feedback since it was present in both practices. The differences and similarities in statements between the two practices answer (i) RQ1. After audio feedback comments, we present results related to visual feedback. The interactions answer all three RQs by showing how participants process visual pronunciation data previously unknown to them. 

The front end results conclude with quantitative findings: the number of times participants recorded with each tool, NASA-TLX, and SUS. Each of the findings help complete the picture painted with qualitative results. 

In general, participants enjoy vowel charts but often do not understand how to interpret visual results or successfully change their pronunciation to affect the visualization. They talked about their vowels more, but did mention other aspects of pronunciation, causing results for RQ2 to be inconclusive. In general, participants engaged with \visowel{} more and appreciated the guidance it gave including the change of a vowel over time in comparison with the audio-only practice due to the reliance on self-perception of audio.

% \todo{latency and stopping recording}
% \todo{discuss research questions here, repeat them again at the beginning of the section}
% \todo{to answer the research question, we must look at... then bring it together}
% \todo{state that part 1 is the usability study, part 2 is the technical result}

% \subsection{Pronunciation Gains}
% There weren't any. Note that a final evaluation is not included for P2 as it was forgotten. We include their qualitative data because we did not have reason to believe that the last step affected how they interacted with the tool. 

% \todo{[insert pronunciation gain table]}

\subsection{User Study: During Practice}
\subsubsection{Auditory Feedback}
\label{sec:AudFeed}
We found four main categories of auditory feedback that participants considered while practicing pronunciation. They brought up similarities to English in words or sounds and connected the sound to how they felt or believed their mouth moved. Linguistic categories, vowels, consonants, and pitch, also impacted participants' understanding of their speech. Finally, they brought up uncertainty regarding the success of pronunciation based on auditory feedback with or without visual feedback.

\paragraph{Similarities to English (3/8)}
Two of the participants used the similarity of Spanish words to English words as a reference for practicing Spanish. P2 focused on the sounds in Spanish in relation to English sounds. "'Loa' wasn't too difficult either. Just like boa constrictor, I think." They did not bring up the relation to English during \visowel{} interaction. P3 and P4 mentioned their pronunciation in relation to English three times during both practices. In both the \visowel{} and audio-only practice they mentioned that they felt some of their pronunciations had a strong American accent. These comments begin to answer RQ1 by suggesting that learners consider L1 speech patterns during interaction regardless of represented visuals of those parts of speech.

\paragraph{Mouth Movement (4/8)}
The connection between mouth movement and subsequent pronunciation was consistent for both conditions. Four of the eight participants brought up how their mouth affected what they were hearing. Half saw the audio-only condition first (N=2). Three participants connected the sound to their mouth movement during interaction with \visowel{} and two mentioned it during interaction with the audio-only feedback, one who saw the audio-only first. Those who mentioned it during interaction with \visowel{} said that they were trying to connect the sound that caused the line to show up on the chart to the tongue position in the mouth (N=3). 
\begin{quote}
    \textit{I'm trying to replicate the position on the chart and think about how the sound is going to be connected to, you know, the tongue position and everything.} (P9)
\end{quote}
Based on these responses, it is possible that seeing a visualization that provides feedback on tongue position brings attention to articulators like the tongue than feedback that does not include that information. 

\paragraph{Aspects of Sound (3/8)}
The participants also discussed different sounds in their speech: vowels, consonants, pitch, and emphasis. Two participants mentioned thinking about consonants, one during interaction with audio-only (P9) and the other during \visowel{} (P7). 

\begin{quote}
    \textit{It kind of sounds like the speaker is saying} [\textsl{\textipa{D5}}] \textit{instead of} [\textipa{\slshape d5}] \textit{and I was saying} [\textipa{\slshape d5}] \textit{instead of} [\textipa{\slshape D5}].~(P7)
\end{quote} 

Without visual feedback, participants considered overall pronunciation such as pitch and emphasis during practice (N=4) while the vowel-specific visualization caused them to focus on the difference vowels (N=3). Only one participant, who saw \visowel{} first, mentioned the vowels during the audio-only portion. Participants compared their pronunciations to that of the Spanish speaker's (N=3) and reflected on modifying their pronunciation based on the Spanish speaker's pronunciation (N=1) with \visowel{}. 
\begin{quote}
    \textit{There's more of an} [\textipa{E}] \textit{to the speaker's sound and there's more of an} [a] \textit{to mine}.~(P8)
\end{quote}
These results begin to answer RQ1 and RQ2 by indicating that a visualization focused on vowels will focus learners on but does not seem to limit their attention to them. On the other hand, participants without visual feedback to guide them did not consider segmentals, other than one who may have been primed by experiencing \visowel{} first.

\paragraph{Uncertainty During Interaction (4/8)}
Four participants mentioned uncertainty during practice with audio-only and/or \visowel{}. While interacting with \visowel{}, participants mentioned feeling uncertain about some aspect of their pronunciation without being able to articulate what was causing their uncertainty (N=3). Most of the comments expressed doubt regarding how the vowel was plotted on the chart in comparison to what they heard.
\begin{quote}
    \textit{I recorded again partially because I want to match my vowel... partially because I want to see how it interprets what I'm saying and see if I agree with that and feel like I'm seeing that.} (P8)
\end{quote}
All four participants mentioned uncertainty when interacting with the audio-only practice. Each of the comments directly related to being unsure why they sounded different to the Spanish recording. In terms of RQ1, it appears that participants can feel equally uncertain about what makes them sound different from a Spanish speaker with and without visuals. Visual information adds another layer of uncertainty-- do they match up with what participants think they heard.

\subsubsection{Visual Feedback}
\label{sec:VisFeed}
All comments regarding visual feedback were made with \visowel{}. Participants expressed opinions on their closeness to the Spanish recording lines and related what they saw appear on the chart with what they felt or heard when they said the word. The results answer the second research question by revealing how participants interpreted results and what they found difficult during interpretation.

\paragraph{Using the Spanish Speaker's Line as a Goal (8/8)}
\visowel{} provided participants with a tangible goal to work toward (N=8), which is in stark contrast with all participants relying on their perception of auditory signals to practice with the audio-only condition (N=8). Three expressed their satisfaction with their speech as getting closer to the Spanish speaker based on the proximity of the Spanish speaker's line. 
\begin{quote}
    That was cool... I didn't expect it to be that much of a linear progression. It really did go from furthest to closer there. (P2)
\end{quote}
Five mentioned that they were surprised by where the vowel showed up on the chart, either because they felt they had not said the vowel well and it showed up closer to the Spanish speaker than expected or because they felt like it had sounded correct and it was not close. In general, they tended to trust the visualization over what they heard. 
\begin{quote}
    \textit{Oh wait... I feel like the "a" was like- it was very much like an} [\textipa{ae}] \textit{sound.} (P4, when a vowel they expected to show up far from the target showed up closer.)
\end{quote}
Learners enjoyed experimenting with the chart when it matched with what they heard and when it seemed like they were improving. Otherwise, they felt discouraged with interaction if they couldn't affect the location of the line or if it was far away from the spa line. They also tended to over rely on the visualization-- if a first recording was close to the Spanish speaker's line, they would not record another time.

\paragraph{Connecting Visual and Auditory Feedback (5/8)}
Difficulties in interpretation of \visowel{} fell into two main categories: the length of the vowel line and the area in which the vowel landed. While some participants ignored the length and shape of lines, others reported uncertainty interpreting them (N=4). Dissatisfaction with the length of the line caused two participants to record again. None of the participants considered how tongue movement affected the lines during speech. 
% , P5 wanted to move the lines while P3 and P7 were not sure if they were able to interact with the chart at all
% P4 and P5 expressed confusion by longer lines and shorter lines, stating that they were not sure why they were long or short. Both indicated a dissatisfaction with the length of the line was a reason for recording again.
% P2 wondered why the vowel chart was not a simple square but instead was represented as a half trapezoid and how the algorithm knew where their tongue was. P7 wanted to know if the colors indicated that they were supposed to record again since one of the self-record words was colored in red for colorblind friendliness.
% A few participants were unsure what interactions were available with \visowel{} (N=3) and
\begin{quote}
    \textit{I wasn't able to make a clear connection between the length of the line drawn and what I was saying. I think it was related to, like, how long I spent on a vowel, but it was hard for me to actually determine how long I was spending.} (P2)
\end{quote}
The location of lines also confused participants, though part of this can be attributed to inaccuracy in formant extraction (N=5). However, participants expressed confusion about location when the algorithm correctly found formants, with one in particular expressing confusion by vowels showing up outside the lines (N=4). 
\begin{quote}
   \textit{In the chart, it says that my tongue is, like, high front. So now, I tried to lower it, but that doesn't seem to be working.} (P4)
\end{quote}
A few participants attempted to adjust their pronunciation in order to move the line (N=3), but only two expressed success in achieving their goal. Both of the participants explicitly used the English corner vowels introduced during calibration to move their tongue in that direction. The success of these participants suggests that including more known vowels during the tutorial phase might help learners associate tongue positions with visuals.
% P6 also checked twice what a calibration word's location was before recording.

\paragraph{Creating a Mental Model (4/8)}
Participants came up with different reasons for why their vowels showed up in different locations. Some thought that the loudness of their speech affected the vowel extractor (N=2). Others thought that their pitch might affect where the vowel showed up with a higher pitch making the line end higher and a lower pitch making the vowel show up lower (N=3). P8 suggested that the calibration or microphone might affect why a vowel showed up in a different place from expected. Finally, P6 thought that speaking English might make vowels show up higher on the chart. It seems that peoples' mental models of speech displayed on graphs may tend toward interpreting increasing and decreasing lines as associated to pitch and amplitude. 

\paragraph{Familiarity}
One of the challenges that participants faced was lack of familiarity with the tongue's role in producing speech (N=2). Pt expressed how this made it more difficult to reach their goals. "I don't think about how I move my tongue, like, ever. So when I'm given instructions ... I don't have the muscle memory of that, like, ideology or specifically how to get there." Although none of the participants said that they had seen a vowel chart before, a few had heard some of the practice words (N=2) and another had seen face-cutaway photos in linguistic content online (N=1). This suggests that interpretation of visuals may not be the greatest barrier in adjusting pronunciation if users do not have previous experience adjusting articulators in accordance with their goals.

\subsubsection{Engagement} 
\begin{table}[]
    \begin{tabular}{l|cc}
    ID & \multicolumn{1}{l}{\visowel{}} & \multicolumn{1}{l}{Audio-Only} \\
    \hline
    P2             & 2.25                              & 1.25                           \\
    P3             & 2.625                             & 1.625                          \\
    P4             & 1.4                               & 1.25                           \\
    P5             & 1.375                             & 1.25                           \\
    P6             & 2.375                             & 1.25                           \\
    P7             & 1.875                             & 2.25                           \\
    P8             & 2                                 & 1.5                            \\
    P9             & 4.25                              & 2.75                           \\
    \hline
    \hline
    Total Avg.     & 2.26875                           & 1.640625                      
    \end{tabular}
    \caption{Average number of recordings per word separated by practice tool across the four words within each practice.}
    \label{tab:engagement}
\end{table}
We define engagement as the number of times participants recorded practice words during \visowel{} and audio-only practice, disregarding a recording if our system did not find a vowel in \visowel{} practice. Because participants stated that they chose to record again when our system did not find a vowel during audio-only practice, we include those recordings. Table \ref{tab:engagement} shows the average number of recordings per word. Most participants recorded with \visowel{} more than with AOF. This is consistent with qualitative results. Every participant found \visowel{} to provide them with actionable feedback that did not depend on their individual judgement. 

\subsubsection{Task Load}
The results for the task load can be seen in Figure \ref{fig:nasascore}. The majority of difference in scores is between the pronunciation practice with the tools and baseline task. It appears that \visowel{} was more of a burden which is complemented by some participants saying that they thought harder to situate the visual with the auditory feedback for their recording and the Spanish speaker's (N=2). 

\subsubsection{System Usability}
\begin{figure}
    \centering
    \begin{subfigure}{\linewidth}
        \centering
        \includegraphics[width=0.9\linewidth]{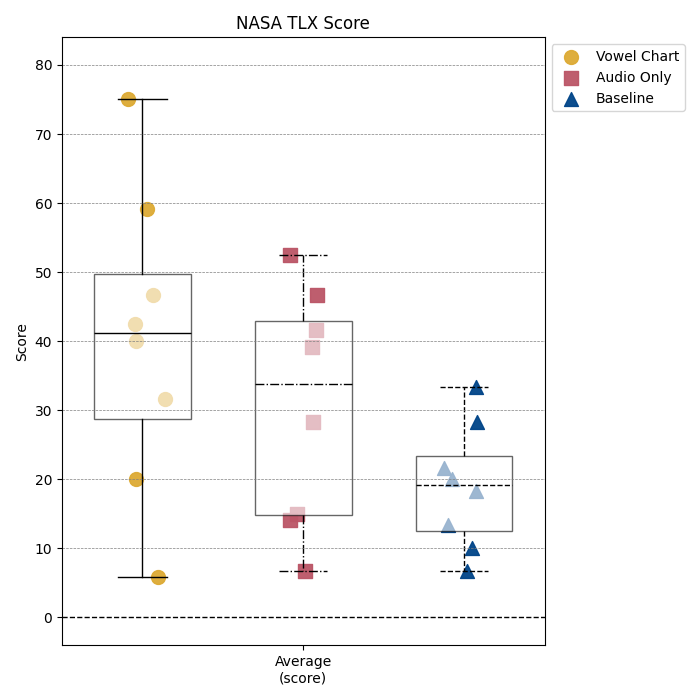}
        \caption{A box and whisker plot of the average NASA-TLX scores for \visowel{}, audio-only, and baseline. The y-axis ranges from 0 to 80.}
        \label{fig:nasascore}
    \end{subfigure}%
    \hspace{1em}
    \begin{subfigure}{\linewidth}
        \centering
        \includegraphics[width=0.9\linewidth]{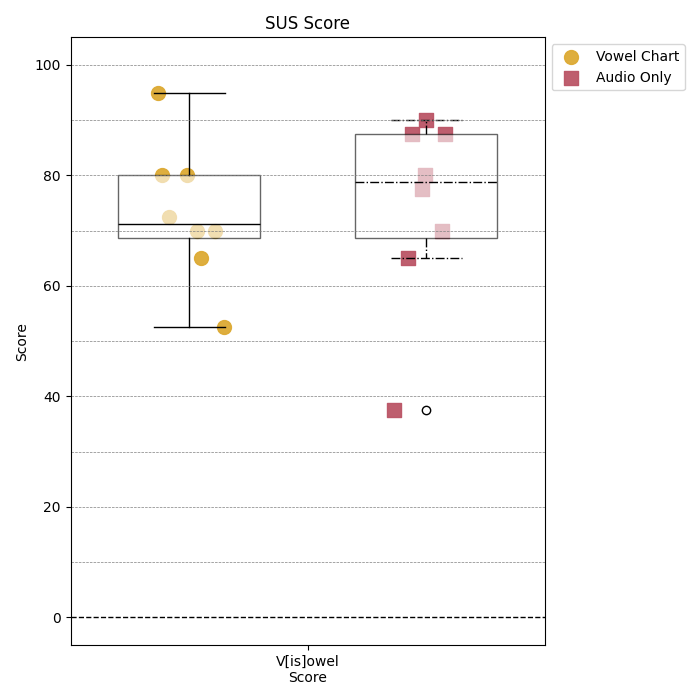}
        \caption{A box and whisker plot of the average SUS for \visowel{} and audio-only. The y-axis goes from 0 to 100.}
        \label{fig:susscore}
    \end{subfigure}
    
    \caption{Average scores from surveys.}
    \label{fig:scooore}
\end{figure}

All but P2 rated both practice systems with an above 65 for system usability (Figure \ref{fig:susscore}). Again, audio-only seems to be easier to use, likely because even if the algorithm did not detect a vowel, an playback box would appear. In \visowel{}, nothing would appear because there were no formants to extract.  

% P2 and P7 connected the Spanish spelling of the word with an English word of the same spelling. 

\subsection{Interview: Reflection on Practice}
\subsubsection{Audio-Only}
In general, participants appreciated the ability to record and listen to themselves as much as they wanted in the audio-only practice, but some wanted to have some sort of feedback instead of completely relying on their own judgment (N=2).  P7 noted that the amount of feedback depended on the word they were practicing, stating that a more complex word would be better with \visowel{}. 

The simplicity afforded participants the ability to listen to themselves and the Spanish speaker as many times as they liked (N=6). P2 felt that the lack of feedback didn't ultimately help them differentiate between pronunciation that was "correct" or "incorrect". On the other hand, P3 preferred the simpler feedback because they felt more confident recording with it. 
\begin{quote}
    \textit{For some reason, I felt like this one I could listen to it more. And so, like, I could focus on listening to it, if that makes sense.} (P3)
\end{quote}

Participants mentioned using the Spanish speaker's audio and their own recently recorded audio to help them hear differences between what they pronounced (N=4). P2 used a trial and error approach to changing their pronunciation based on the difference in emphasis, while P6 and P7 spoke generally about hearing what was wrong in their recording based on what they heard in the Spanish recording. P9 reflected on how the audio helped them move away from how English letters did not have the same pronunciation of Spanish letters.

\begin{quote}
    \textit{\visowel{} help me go forward and understand how exactly to pronounce different vowels in different situations.. It'll help me with the pronunciation in the future. Just thinking more carefully about that.} (P9)
\end{quote} 

Most of the participants felt confident about their pronunciation (N=5), though some mentioned that their confidence could be misplaced since it was based on their own judgment (N=2). While P4 expressed it as a leap of faith, P3 appreciated the lack of visual feedback because, "I wasn't seeing how I was getting farther and farther away... I feel like I could match myself to the speaker more, if that makes sense." Two participants said that they were confident, but some words were just hard for them to pronounce because they did not have a Spanish accent. The rest of the participants did not feel confident interacting with only auditory feedback (N=3). Two stated this was because they were just guessing. P9 explained further that, "having an untrained ear is just kind of hard to tell exactly what I did." 

\subsubsection{\visowel{}}
Participants found \visowel{} interesting, though not always accurate (N=3). P8 assumed that something was off with the algorithm while P3 and P7 attributed differences in their pronunciation to lack of experience with the chart and Spanish respectively. On the other hand, three participants preferred the visual feedback to audio-only because it gave them tangible information to act on, either as a visual goal (N=2), or feedback on the position of the tongue (N=1). Participants appreciated having feedback that directly related to a physical space since that gave them a direct path to change how they were speaking (N=2), while others expressed appreciation for a visualization that showed the progression of vowel over time since a vowel sound could start off in the right space but end somewhere different from the goal (N=3). 
\begin{quote}
    \textit{I thought it was interesting to kind of see, especially, the transitions between certain parts of sounds in the vowel.} (P8)
\end{quote}

Two participants noted that they liked seeing what they might be doing wrong in the context of thinking that they had done a good job (N=2).

While participants appreciated the ability to work toward a goal, they were not always sure how to interpret what they were seeing (N=2). P6 ignored the length of the line altogether and focused on getting their lines to be in the same region of the Spanish speaker. P5 noted that although it would take longer to understand \visowel{}, they found it more useful for improving pronunciation. P9 noted that \visowel{} not only helped them practice their phonetics, but also helped them focus on which part of the word was stressed by the Spanish speaker. 

All but one participant felt confident about adjusting their pronunciation with \visowel{} (N=7), but attributed their confidence to different aspects. P5 said that they were comfortable adjusting their pronunciation, but noted that they had to think about the visual and auditory information so it took longer. Four said they were confident adjusting their pronunciation based on the feedback because they could see their progress, while one stated that they did not think the chart always accurately displayed what was wrong.  
\begin{quote}
    \textit{I felt confident in my ability to change the line a little bit, but not in a way that I would get an answer that was necessarily more accurate.} (P8)
\end{quote}

In general, participants expressed hesitancy regarding the accuracy of the vowel chart (N=6). Half explicitly stated that they thought it might not being giving accurate feedback, while the rest attributed the feedback to a difficulty on their part understanding how to move their tongue to get the lines to change on the chart.

\subsection{Technical Evaluation}
% Engineering evaluation
% Describe the critical components that *must* work for you to use it in the real world
For our tool to work long term in the real world, the formant extraction must choose the correct frequencies for $f_1$ and $f_2$ and extract vowel onset and offset times within a set tolerance. 

Any major mistakes during formant extraction will result in inaccurate results for the position of the tongue as displayed on \visowel{}, creating more confusion for the user and potentially negatively impacting their pronunciation. Slight variation for a formant is considered a small error. We report the mean absolute error (MAE) in such cases. If the algorithm picks the wrong frequency for a formant, that is categorized as a large error. We report the probability of such an error occurring. 

% Since this is a real-time system, we present the latency from when a user stops speaking for recording and when a vowel appears on the chart for \visowel{} or when a recording box appears for audio-only. As our recording system automatically stops recording when it stops hearing a voice, we also report the likelihood of it inappropriately stopping recording.

We calculate the mean absolute error (MAE) for vowel onset and offsets. We use the MAE to compare our algorithmic results with manually extracted formant values by Praat. The Praat algorithm computes the formant values within the whole signal by the Burg algorithm~\cite{hugoQuenePraat}. By selecting the vowel, we outputted a list of formants and timestamps for the vowel.

\subsubsection{Vowel Boundary Accuracy}
\begin{table}[t]
    \centering
    \begin{tabular}{l|ccccccc}
    \hline
        ~ & Word & MAE Onset (ms) & MAE Offset (ms)\\ \hline
        P2 & pie & 5 & 15 \\ 
        ~ & tara & 5 & \textbf{25} \\ 
        P3 & canoa & \textbf{25} & \textbf{265} \\ 
        ~ & duda & \textbf{15} & 5 \\ 
        P4 & polo & \textbf{15} & \textbf{15} \\ 
        ~ & talla & 5 & 5 \\ 
        P5 & pe & 5 & 6 \\ 
        ~ & cana & 5 & \textbf{135} \\ 
        P6 & oreo & 5 & \textbf{155} \\ 
        ~ & pie & 5 & 5 \\ 
        P7 & talla & 5 & 5 \\ 
        ~ & poleo & 5 & \textbf{205} \\ 
        P8 & oro & 5 & \textbf{125} \\ 
        ~ & duda & 5 & 5 \\ 
        P9 & oro & 5 & \textbf{85} \\ 
        ~ & cana & 5 & \textbf{145} \\ 
    \end{tabular}
    \caption{MAE for a randomly sampled set of words. Bold indicates an offset that is outside allowed tolerance.}
    \label{tab:vwlBounds}
\end{table}
We randomly selected audio files to hand calculate the onset and offset times for the vowel. MAE for vowel boundaries was generally within tolerance (<10ms) for onset times but was completely out of tolerance for offset (Table \ref{tab:vwlBounds}), which increases the likelihood that participants saw accurate formants for the beginning of a vowel and inaccurate for the end. 

\subsubsection{Formant Extraction Accuracy}
\begin{table}[t]
    \begin{tabular}{l|cc}
    ID & MAE $f_1$ (Hz) & MAE $f_2$ (Hz) \\
    \hline
    P2              & 43.44                     & 83.82                        \\
    P3              & 41.26                     & 94.15                        \\
    P4              & 47.19                     & \textbf{109.69}                       \\
    P5              & 65.60                     & 95.51                        \\
    P6              & 80.08                     & \textbf{122.24}                       \\
    P7              & 53.99                     & \textbf{125.74}                       \\
    P8              & 84.19                     & \textbf{182.78}                       \\
    \hline
    avg. MAE        & 59.39                     & 116.2757143
    \end{tabular}
    \caption{The average Mean Absolute Error (MAE) of $f_1$ and $f_2$ for each participant. Bold indicates an offset that is outside allowed tolerance.}
    \label{tab:MAEformant}
\end{table}
% \footnotetext{}
% \label{footnote:excludeMAE}

On average, MAE of the formant frequency was within the threshold for acceptable difference of 100Hz for $f_1$ (Table \ref{tab:MAEformant}). The formant extraction for $f_2$ was not as accurate, only being under the threshold for a few participants (N=3), which is expected given the low accuracy for vowel offsets. The likelihood of a large error varied significantly between participants, ranging from 5\% to 26\% (Table \ref{tab:formantLarge}). Because of the high likelihood of $f_2$ inaccuracies, participants often saw visually inconsistent results between recordings. One recording would show up in the appropriate quadrant of the chart but a similar production's recording could show up on the opposite side. 

\begin{table}[t]
    \centering
    \begin{tabular}{l|ccc}
        ID & \# of Large Errors & Total Recordings & \% Incorrect  \\ 
        \hline
        P2 & 4 & 28 & 14.29  \\ 
        P3 & 3 & 32 & 9.38 \\ 
        P4 & 3 & 29 & 10.34 \\ 
        P5 & 1 & 21 & 4.76 \\ 
        P6 & 7 & 27 & 25.93 \\ 
        P7 & 8 & 32 & 25.00  \\ 
        P8 & 4 & 24 & 16.67  \\ 
    \end{tabular}
    \caption{The percentage of large formant errors for each participant, where large is getting the formant completely wrong.}
    \label{tab:formantLarge}
\end{table}

% \todo{include \# of participants per statement and add
% paragraph assertions of what the para is about}

\subsection{RQ1: How does interaction between a visual and audio-only feedback method differ?}
Despite mistakes made by our formant extractor and confusion in interpreting what the lines on the chart meant, participants still chose to practice with \visowel{} more than audio-only feedback (AOF). While some of this could be due to the novelty effect, based on verbal feedback during interaction, participants were also motivated by producing speech that would get the vowel just a little bit closer to the Spanish speaker's. They chose to practice with the vowel chart because \textbf{they had guidance on what to change}, the visual giving them a basis to adjust their pronunciation beyond auditory perception and a way to evaluate after recording. Despite showing a desire to engage with \visowel{}, participants expressed a degree of uncertainty that did not exist with AOF. Many participants said that they felt confident practicing with audio-only feedback, while recognizing that the confidence was based on their own perception. Confidence was lower during interaction with \visowel{} because of the conflict between visual and auditory interpretation. The conflicting information could contribute to the higher average recordings made with \visowel{} since participants were trying to bring the audio and visual feedback into agreement. 

Neither method of feedback affected whether participants thought about other aspects of speech, which indicates that despite placing an emphasis on visual feedback, \visowel{} may not cause a blinder effect. Just under half of the participants noted that they liked seeing the progression of the vowel over time (N=3), suggesting that our novel approach to vowel visualization is a desired feature. In short, \visowel{} was treated with some distrust, but that did not cause participants to dislike it or to practice with it less. 

\subsection{RQ2: How do phonetically untrained users interpret vowel charts during interaction?}
Participants liked the target Spanish lines on the chart because it gave them a tangible goal to work toward. We noticed that all participants were not sure how to interpret the length of the lines, how to affect the length, or what kinds of interactions were available. We believe that there are three aspects that led to participants' confusion based on the qualitative and quantitative data. First, participants were not used to thinking about how their tongues moved, which would hinder recognition of the tongue's movement during the vowel. Second, the algorithm picked up too much of the surrounding consonants. Finally, although the \visowel{} tutorial included a description that discussed what the line length meant, it did not allow participants to experiment with line length.

Many participants thought of where their tongue's position based on visual feedback, despite being unaccustomed. Not everyone felt successful adjusting their tongue position, but those who were used known sounds to move their tongue in that direction. As a result, we believe that having known target vowels is important for success during initial interaction with \visowel{} since the tongue will know what position to take for the known vowel sound, giving it a direction to move. 
%We believe that a short explanation on basics of what the backend is doing and more examples in the tutorial would help participants create a better mental model of how to effectively implement the feedback. 

More practice in the tutorial would provide learners with more known words on the chart, leading to a better understanding of how to move the tongue to get the desired change in location and reduce the possibility of misunderstanding which axes map to different parts of the tongue's position. For example, a few participants mentioned the calibration words to remind them where known words landed. But they were not always correct, indicating a conceptual error~\cite{booth1991errors}. The strength of the vowel chart lies in its mapping to the physical realm. If participants are confused by what that mapping is, that conflicts with their ability to implement feedback. Including examples in the tutorial that provide learners with more points of reference, they will have a better chance of knowing how to get their line to move across the chart. 
% An example might ask them to produce vowels that move from the back to the front of the chart, instead of asking for words with vowels in them.   
% One participant misplaced the upper right vowel's location as being in the lower right. Another participant said that they tried to lower their tongue when they wanted the line to move higher on the chart.

Participants often thought that pitch might affect what they saw on the chart. We did not bring up any of these in our introduction to \visowel{}, which implies that people might naturally attach pitch to lines that vary in height on a chart. One participant incorrectly concluded that the origin of language would affect the chart, which is not the case because of our method of calibration. A short exercise in the tutorial focusing on pitch should alleviate misunderstandings.

Due to our algorithm, participants saw mistakenly plotted vowels. They put greater emphasis on the visualization than on their perception of pronunciation, which lines up with statements made about how discerning if their pronunciation was correct based only on audio feedback sometimes was just guessing. Due to the trust put in the visualization, it is important that there are minimal mistakes made by the algorithm. 

In conclusion, users found the target based method of a vowel chart to help them practice pronunciation. They struggled to understand what made vowel lines long or short and attributed changes in the angle of the line to a change in their pitch during production. 
% Interaction with \visowel{} caused curiosity as well, since multiple participants questioned how \visowel{} worked and conjectured on what affected the plotted lines.   
\subsection{RQ3: How does a visualization that focuses on vowels affect users' perceptions of other aspects of pronunciation?}
A concern with practicing on a specific sound in a language would be that it creates a blinder effect toward other important aspects of pronunciation. Based on comments made during interaction, it is not clear whether this happens or not. Multiple participants mentioned a desire to improve their pitch, But only two participants thought about consonants, one during interaction with \visowel{}, the other during audio-only after interaction with \visowel{}. It appears that participants continue to consider multiple aspects of pronunciation based on their comments, but more research is needed to confirm these preliminary findings. 
% Possible perceptual challenges
% \todo{reflect on the results of the whole. Would I use this? pull the results together}
% \todo{talk about the difference with consonants, talk about what i would have done differently}
%Describe next steps
\section{Discussion}
\paragraph{Real-world use.} We note that \visowel{} is a vowel-specific interactive tool. It still struggles with certain consonants which can limit its real-world use. Future work may address this limitation by using additional linguistic properties in conjunction with our methods. Additionally, users could also be given greater freedom on which vowel should be targeted for practice.

Any incorrect feedback will lead users astray, which is likely to be detrimental over an extended period of time. Based on offset times, \visowel{} currently gives faulty feedback. Another possible effect of using \visowel{} for a longer time is that it could put undue emphasis on vowels. While initial interactions are encouraging since participants continued to think about pronunciation as a whole, there were many more comments regarding vowels. 

\paragraph{Longer tutorial.} 
We noticed a tendency for participants to trust the visual feedback over their own ears. Because there is no statistical way to capture just the vowel and exclude all of a consonant, introducing learners to the concept of sonorant consonants should help reduce confusion regarding longer lines on the chart. To further reduce excess visual information, we could simplify vowel lines, potentially allowing toggle between an average location and the whole vowel.

Given two participants success with anchoring vowels based on English vowels, we believe that including more information during practice explaining how to interact with \visowel{} could help. This might include more English words with different vowels to provide more reference points, and feedback to emphasize the physical connection to the chart by giving relational feedback, e.g. "Your vowel is lower than the goal, try bringing your tongue up to get closer". 

% Participants also mentioned that the spelling of some words reminded them of English sounds. We will hide the words initially so that participants can focus on hearing the sounds and not see conflicting visual information.

One of the takeaways from participant responses to \visowel{} is the need for greater transparency on how the system extracts their vowels. We received multiple comments regarding the confusion of what affected the position on the chart, one being that participants thought the pitch of their voice might make the line show up in different places. The tutorial could have practice that contrasts loud production with soft and high with low pitched speech.
% Participants were also unsure what caused the system to be unable to catch their vowels, with some assuming it might be because they were speaking too loudly. 

% Participants expressed other aspects of the practice that were difficult for them. P2 said they found the vowel chart the most confusing part of practice since they weren't aware that they could click on a line and see where the line began and ended. They suggested putting an arrow on one end to indicate the end of the vowel. \visowel{} was also confusing because it was hard to understand how the tongue moves as it produces speech. Two participants found the inclusion of resonant consonants on the chart as confusing. P3 added that they found the extra lines distracting and were unable to get their speech to different areas of the chart. Similarly, P7 said they often were not sure what to do in either practice scenario and "went off vibes".

\paragraph{Longitudinal study.}
Our goal is to improve intelligibility in a second language, not attain native-likeness. Our visualization, though an improvement over frequency markings and addressing previous limitations, does not provide visual markers sufficient for beginner learners. Based on the feedback, and to de-emphasize native-like pronunciation, incorporating an intelligibility threshold would be helpful for a second language learners. 

Another limitation to our findings is the novelty effect; none of the participants had seen the visualization before. However, since participants mentioned using the Spanish speaker's visual as a goal better explain why participants greater engagement with \visowel{}. A study over a longer period of time would allow us to understand how much of the engagement was due to novelty.

\paragraph{Applicability for other languages.}
\visowel{} can be extended to other languages, provided that they primarily differ from the first language in tongue location. We could add to the current visuals with additional language specific contexts, such as differences in lip rounding or nasalization. For future work, we could create a catalog of language specific idiosyncrasies, which could be loaded based on user needs.
\section{Conclusion}
State-of-the-art CAPT tools do not provide personalized audio and video feedback for pronunciation improvement. We find that personalized feedback can have a significant impact on a learners' pronunciation in a second language. To that effect, we introduced \visowel{}, an interactive vowel chart that provides visual and auditory feedback based on a learner's speech patterns. 
Our system creates a personalized vowel chart by extracting their L1 vowel space in a calibration phase. During pronunciation practice, our algorithm extracts the vowels from the recorded words, and plots the formants in the vowel chart, along with the associated audio for feedback, allowing for learners to interpret their pronunciation based on both visual and audio feedback. Findings from our user study suggest that our system supports engaging and personalized learning experiences. Future work includes expanding to more attributes of vowels and other languages.

\begin{acks}
Many thanks to members in the Social Spaces group for their unfailing suggestions and encouragement. Thanks to Shreyansh Agrawal who helped code the front-end of recording audio and Nicholas Kiesel for debugging the back end more times than we can count.
\end{acks}

\bibliographystyle{ACM-Reference-Format}
\bibliography{general}
% TC:endignore
%%
%% If your work has an appendix, this is the place to put it.
% TC:ignore
% \newpage
\appendix
\section{Statistical Analysis}
\subsection{Engagement}
A Q-Q visual test revealed that the number of times participants recorded is not normally distributed~\cite{das2016brief}. Since the samples are not independent, we used the Friedman test to determine statistical significance~\cite{sheldon1996use}. We found that participants recorded more with \visowel{} practice than with audio-only (p=0.0339). Because the visualization in \visowel{} depended on speaker, we tested for a difference between participants who practiced with a female voice (N=4) versus a male voice and did not find a statistically significant difference (p=1).
\subsection{NASA-TLX}
In order to determine the suitability of performing an ANOVA on average NASA scores~\cite{st1989analysis}, we used visual Q-Q graphs to see how much our distribution varied from a normal distribution. We compared the visual results with a Shapiro-Wilk numerical test and determined that the data could be considered to be pulled from a normal distribution (Figure \ref{fig:nasascore}). We performed a two-factor ANOVA without Replication on the average NASA TLX scores collected after the baseline check, audio-only, and \visowel{} practices (p=0.008). We excluded the baseline scores in a second ANOVA and found that there was no statistical difference between the load of \visowel{} and audio-only practices (p=0.102), which is expected since the baseline condition does not require thinking about pronunciation.
\subsection{System Usability Scores}
A test for normalcy in the distributions showed that the \visowel{} scores were close to a normal distribution, but the audio-only was not. As a result, we used the Friedman test to determine if the difference in usability between the tools was statistically significant. There was no statistical difference in usability (p=0.480).

\section{Motivation}
To better understand learning goals, we had participants complete a motivation survey, modeled after the mini-AMTB by Tennant and Gardner~\cite{tennant2004computerized}. The survey gauges the motivation behind learning Spanish using five point Likert scales. 

The overall motivation of a participant as defined by Tennant and Gardner is:
\begin{align*}
    MOTIV = MI + D + AL\label{eq:motiv}
\end{align*}
% \[MOTIV = MI + D + AL\]\label{eq:motiv}

where MI is the numerical response to the motivation question, D is the response to the desire question, and AL is the attitude toward learning the target language (see Appendix \ref{app:motiv} for a list of questions). To return to a 5-pt Likert scale, we divide the result by 3.
\subsection{Questions}
\label{app:motiv}
\paragraph{Integrative Orientation (IO)} If I were to rate my feelings about learning Spanish in order to interact with Hispanic or Latino speakers, I would have to say they are, "weak ... strong" 
\paragraph{Attitude toward Hispanic or Latino Americans (AFA)} My attitude toward Hispanic or Latino speakers is, "unfavorable ... favorable" 
\paragraph{Interest in Foreign Languages (IFL)} My interest in languages other than Spanish and English is, "very low ... very high"
\paragraph{Desire to learn Spanish (D)} My desire to learn Spanish is: "weak ... strong"
\paragraph{Attitude toward learning Spanish (AL)} My attitude toward learning Spanish is: "unfavorable ... favorable"
\paragraph{Instrumental Orientation (IO)} If I were to rate my feelings about learning Spanish for practical purposes such as to improve my occupational opportunities, I would have to say they are "weak ... strong"
\paragraph{Motivational Intensity (MI)} I would characterize how hard I work at learning Spanish as "very little ... very much"
\section{Results}
\begin{table}[]
    \begin{tabular}{l|ccc||c}
    ID & D & \makecell{MI} & \makecell{ALSpa} & Average MOTIV \\
    \hline
    P2             & 4                           & 3                           & 5                                        & 4.0           \\
    P3             & 5                           & 3                           & 5                                        & 4.0           \\
    P4             & 5                           & 3                           & 5                                        & 3.7           \\
    P5             & 5                           & 5                           & 5                                        & 5.0           \\
    P6             & 4                           & 3                           & 5                                        & 3.7           \\
    P7             & 4                           & 4                           & 5                                        & 4.0           \\
    P8             & 4                           & 2                           & 5                                        & 3.3           \\
    P9             & 4                           & 2                           & 5                                        & 3.3          
    \end{tabular}
    \caption{Responses to the motivation questionnaire, where average motivation is calculated as
    ( Desire + Motivational Intensity + Attitude toward Learning Spanish ) / 3}
    \label{tab:motiv}
\end{table}

The participants all had a greater than 3 average motivation to learn Spanish (Table \ref{tab:motiv}). The motivation to engage with Spanish practice is reinforced by the thoughtful comments participants made during practice (Sections \ref{sec:AudFeed} and \ref{sec:VisFeed} ).

\section{Exit Interview Questions}
Each question was asked for both practices.
\begin{itemize}
    \item How did you like the feedback?
    \item How was the feedback useful? 
    \begin{itemize}
        \item What feedback could you act on?
    \end{itemize}
    \item How trustworthy was the feedback?
    \item What did you find confusing or distracting?
\end{itemize}

\section{Design Decisions}
\label{app:design}
\subsection{Color Palettes}
\label{app:palette}
\begin{figure}[h]
    \centering
    \includegraphics[width=0.5\textwidth]{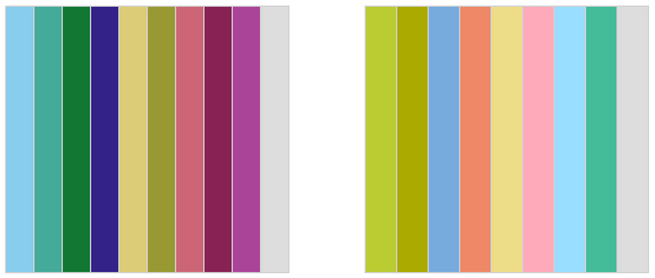}
    \caption{Two colorblind friendly palettes. From left to right: tol\_muted and tol\_light.}
    \label{fig:colorblind}
\end{figure}
We considered colorblind friendly colors for differentiating the words and speakers, green and purple for Spanish, and blue and red for the learner (Figure \ref{fig:visowelPrac}). 

% Since we wanted participants to interact with the Spanish lines on the chart, we initially plotted both Spanish recordings in orange. Participants expressed confusion during interaction, so we gave the words different colors. 

\subsection{Layout}
\label{app:layout}
\begin{figure*}[t]
    \centering
    \includegraphics[width=\linewidth]{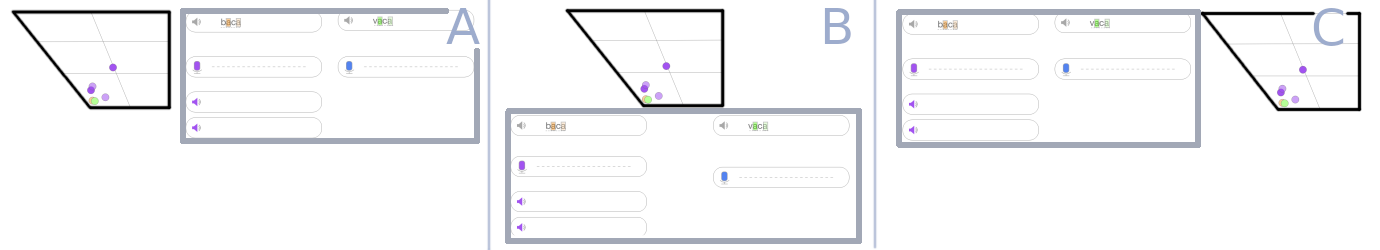}
    \caption{Three potential layouts for \visowel{} practice. The vowel chart is outlined in black and the recording buttons and recordings are outlined in grey. Audio-only practice is just the grey portion.}
    \label{fig:layouts}
\end{figure*}
We evaluated three potential layouts for the minimal pair recording buttons and \visowel{} (Figure \ref{fig:layouts}). The layouts represented the potential orderings that were possible given the chart and buttons. We ended with a horizontal alignment of the buttons with the chart because the vowel chart has a straight side which makes the best use of the available space. 

\section{Tutorial}
\label{app:tutorial}
\begin{figure}[h]
    \centering
    \begin{subfigure}{\linewidth}
        \centering
        \includegraphics[width=0.9\linewidth]{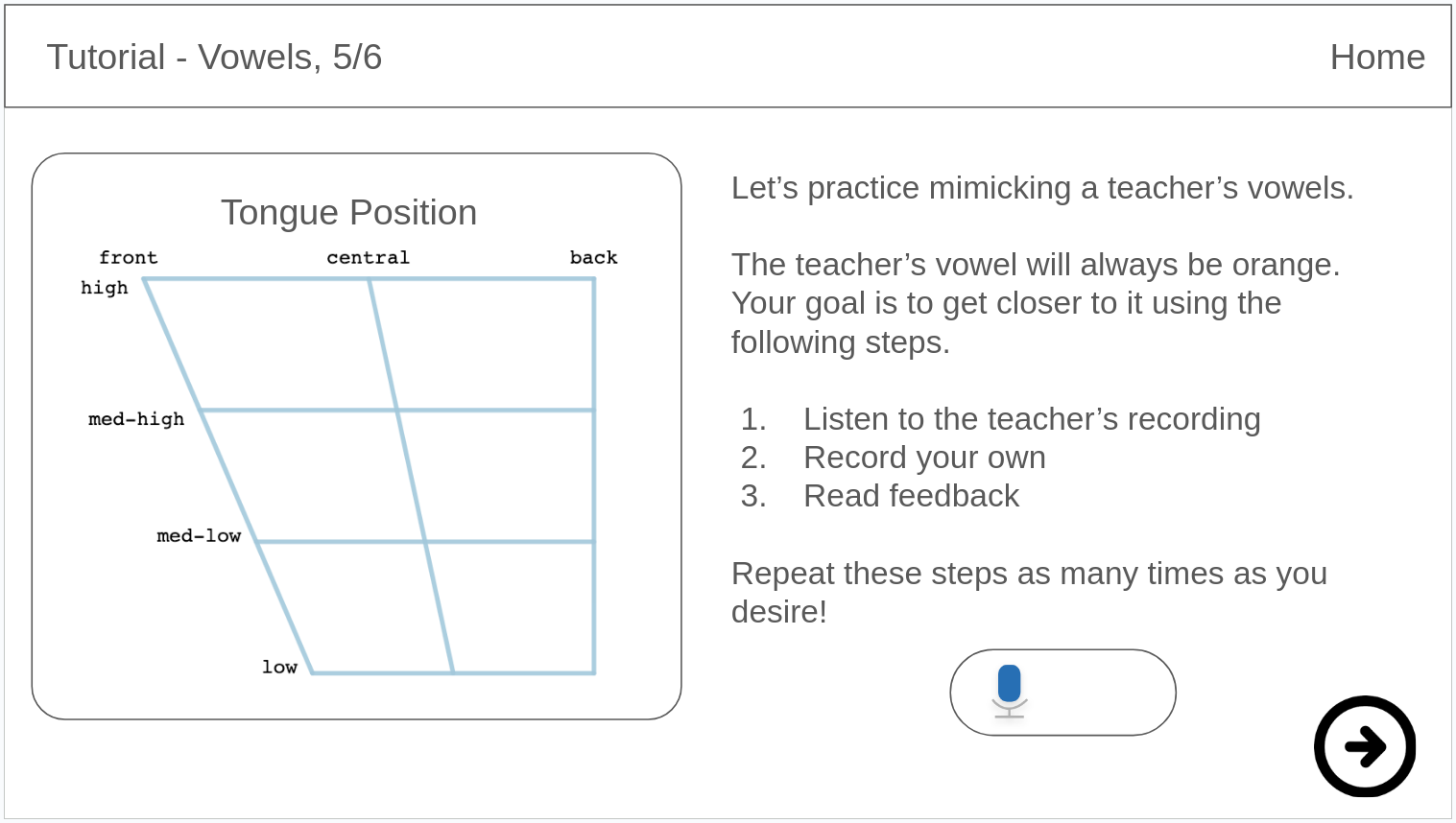}
        \caption{Paper prototype of tutorial page 5. The user is introduced to the idea of the Spanish speaker's line by recording the Spanish word.}
        \label{fig:5tutPaper}
    \end{subfigure}
 
    \begin{subfigure}{\linewidth}
        \centering
        \includegraphics[width=0.825\linewidth]{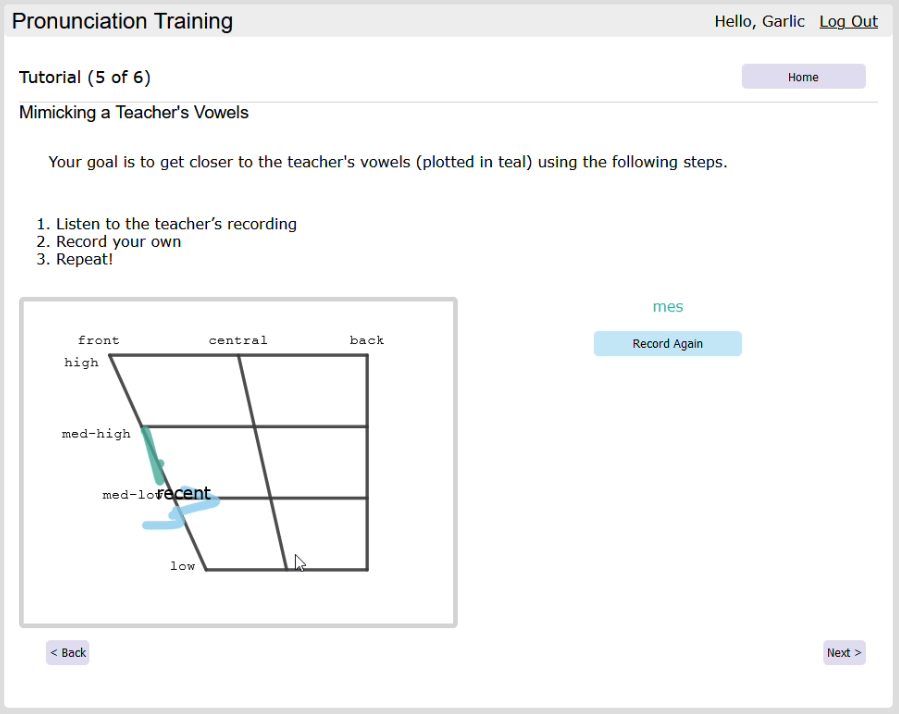}
        \caption{Final interface for the tutorial's fifth page. The user can click on the Spanish speaker's line to hear the word and record their own version of it.}
        \label{fig:5tut}
    \end{subfigure}
    
    \caption{Three pages representative pages taken from the tutorial. The images on the right are the initial paper prototype and the ones on the left are the final rendition of the corresponding pages.}
    \label{fig:tutorialPages}
\end{figure}
\begin{figure}[h]
    \centering
    \begin{subfigure}{\linewidth}
        \centering
        \includegraphics[width=0.9\linewidth]{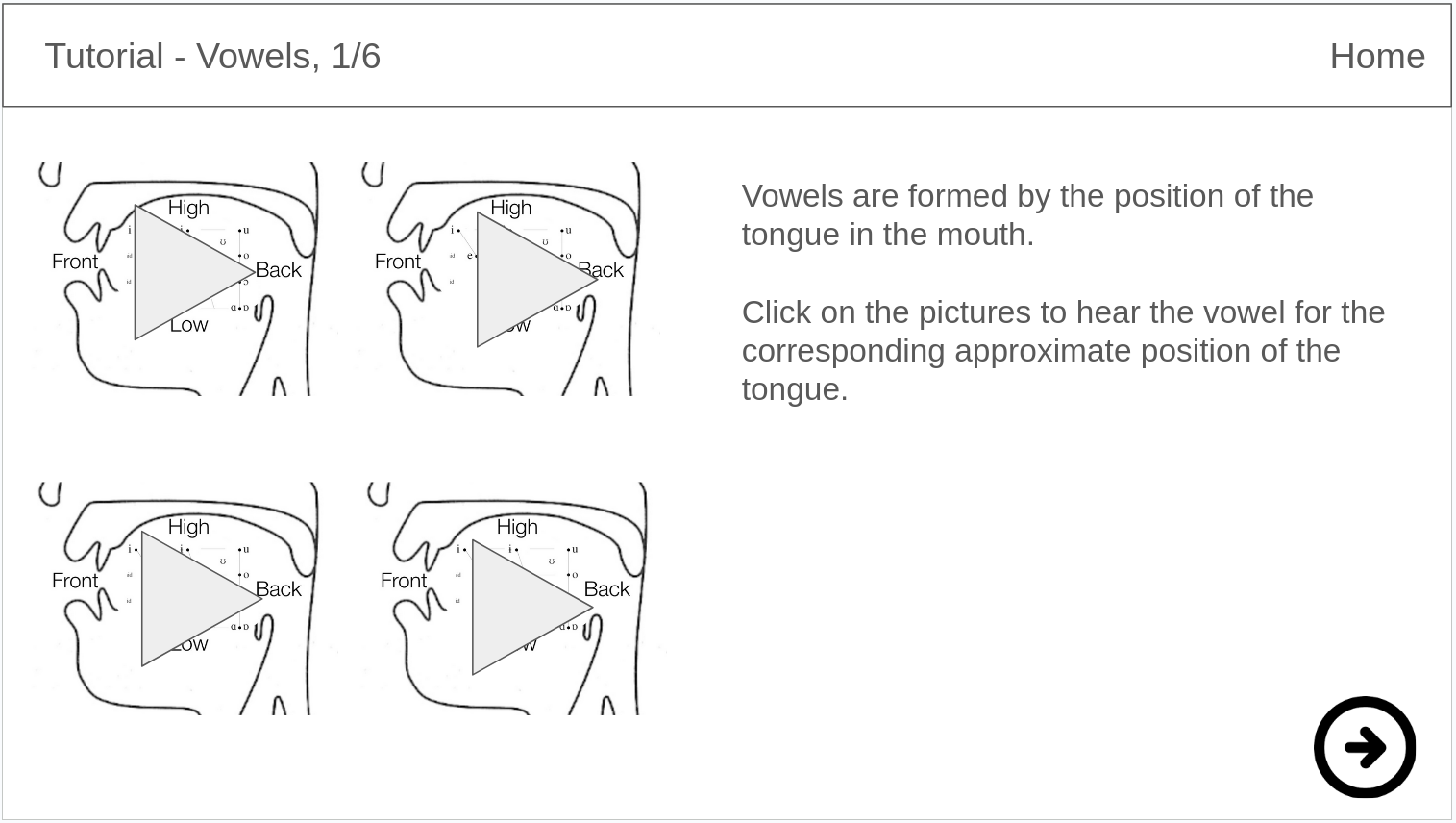}
        \caption{The paper prototype of the first page of the tutorial. A grid of four face cutaways is on the left. Triangles on each of the pictures indicated that they were short videos.}
        \label{fig:1tutPaper}
    \end{subfigure}%

    \begin{subfigure}{\linewidth}
        \centering
        \includegraphics[width=0.8\linewidth]{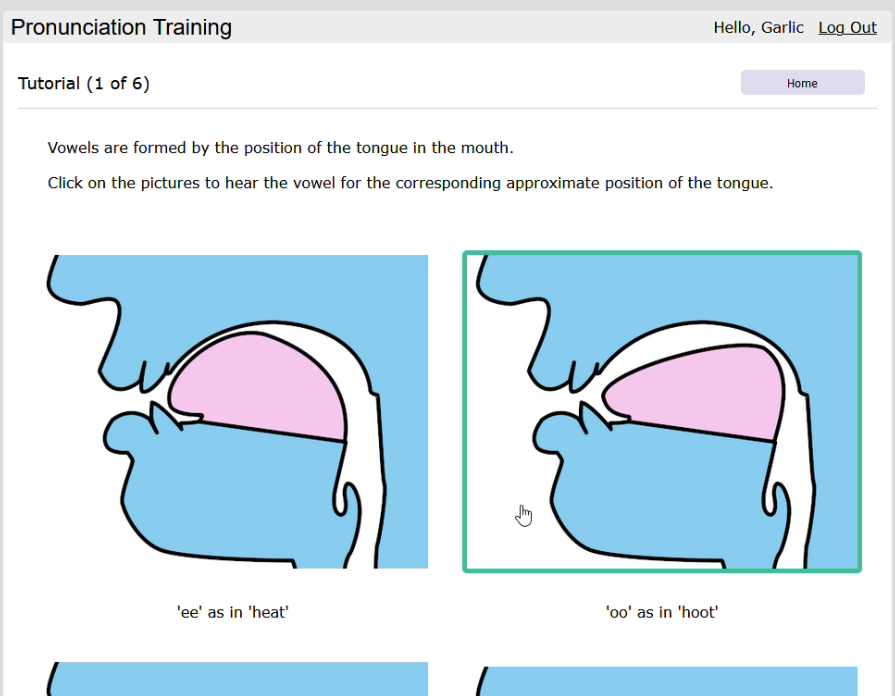}
        \caption{Final interface for the tutorial's first page. The face cutaways are static due technical limitations. Hovering over a picture lines it with green indicating that a user can interact with it.}
        \label{fig:1tut}
    \end{subfigure}
    \label{fig:pg1}
    \caption{Design cycle of page 1}
\end{figure}

The tutorial begins with face-cutaways that show the position of the tongue during the production of four different vowels in extreme positions (two front vowels, high and low, and two back vowels, high and low. See Figure \ref{fig:1tutPaper}). The second page we introduce the vowel chart alongside a face-cutaway to show correlation between the position of the tongue and the points on the chart. The third and fourth steps break down the horizontal and vertical axes by bringing attention to the position of the tongue to a user and letting them record English words that differ in the respective axis. After allowing users to interact with English words, we presented them with the idea of getting close to a Spanish recording. The final step was a review of the tongue's mapping to the vowel chart.  
Based on pretests, we modified the tutorial to include the same four vowels to cement the association from the first step (Figure \ref{fig:2tut}). We created an interactive prototype and ran another pretest. We learned that people were unaware that the face cutaways on the first page were interactive. When clicked, they played the associated vowel. To bring attention to the interaction, we show an outline when the mouse hovers over any of the pictures (Figure \ref{fig:1tut}). Originally, all the vowels on the second page were the same color. Based on feedback, we colored the tongues to match the colors of the vowels on the chart to create a visual association. An overall modification we made based on pretest information was to put the instructions above the vowel chart to encourage users to read the instructions before plowing ahead. We also pared down the instructions to be more concise.

\section{Vowel Boundary Algorithms}
\label{app:algo}
We outline brief descriptions of the algorithms we tested for vowel boundary extraction and why we did not use them. 

We started using a DTW algorithm~\cite{bellman1959dtw}, which maps similar points between two audio samples regardless of speed. There were two flaws with DTW. First, if an English speaker didn't pronounce the consonants like the Spanish reference, then the mapping would include multiple points near the beginning of the vowel, potentially throwing off the beginning and ending boundaries for the vowel (Figure \ref{fig:DTW}). Second, even when DTW was accurate, the window size could not be made small enough (<10ms) to exclude enough of the consonant in the extraction. 

Next, we tested extracting timestamps by using pocketSphinx (PS)~\cite{huggins2006pocketsphinx}. PS is a speaker-independent continuous speech-recognition algorithm begun by Huggins-Daines et. al. It takes in an audio file and returns the phones and silences with associated timestamps detected in the file. The timestamps had a similar degree of inaccuracy as DTW.

We decided to test Praat's vowel extractor suitability for extraction using the parselmouth library, a python interface for Praat~\cite{jadoul2018introducing}. The original Praat script written by Hugo Quené only extracts a portion of the vowel so we modified the script to output the timestamps for the whole vowel~\cite{hugoQuenePraat}. The boundaries using this method were also inaccurate. 

\section{Timing Considerations}
% Timing considerations
We wanted to keep the timing between recording and visualization low. Some parts of the server side code have a set amount of time: speech recognition, vowel boundary and formant extraction. While the algorithm ran these processes, we displayed "Processing" on the record button and disabled interaction with it. For the formants to be displayed on the chart, they must be transformed from the frequency to svg domain. Initially, we transformed each pair of formants, $f_1$ and $f_2$, one by one. We optimized the process by transforming formants in a batch. 

% TC:endignore
\end{document}